\documentclass[twocolumn]{aastex61}
\usepackage{color}

\newcommand{\vrad}{{$v_{\mathrm{r}}$}~}
\newcommand{\kms}{{km s$^{-1}$}~}
\newcommand{\kmsend}{{km s$^{-1}$}}

\received{2017 October 27}
\revised{2018 April 24}
\accepted{2018 April 25}
\shorttitle{Spectroscopy of M85 GCs}
\shortauthors{Ko et al.}
\begin{document}

\title{Gemini/GMOS Spectroscopy of Globular Clusters in the Merger Remnant Galaxy M85}

\correspondingauthor{Myung Gyoon Lee}
\email{mglee@astro.snu.ac.kr}

\author{Youkyung Ko}
\affil{Astronomy Program, Department of Physics and Astronomy, Seoul National University, 1 Gwanak-ro, Gwanak-gu, Seoul 08826, Republic of Korea}
\affiliation{Department of Astronomy, Peking University, Beijing 100871, People's Republic of China}
\affiliation{Kavli Institute for Astronomy and Astrophysics, Peking University, Beijing 100871, People's Republic of China}

\author{Myung Gyoon Lee}
\affiliation{Astronomy Program, Department of Physics and Astronomy, Seoul National University, 1 Gwanak-ro, Gwanak-gu, Seoul 08826, Republic of Korea}

\author{Hong Soo Park}
\affiliation{Korea Astronomy and Space Science Institute, 776 Daedeokdae-Ro, Yuseong-Gu, Daejeon 34055, Republic Korea}

\author{Jubee Sohn}
\affiliation{Smithsonian Astrophysical Observatory, 60 Garden Street, Cambridge, 02138, USA}

\author{Sungsoon Lim}
\affiliation{Department of Astronomy, Peking University, Beijing 100871, People's Republic of China}
\affiliation{Kavli Institute for Astronomy and Astrophysics, Peking University, Beijing 100871, People's Republic of China}

\author{Narae Hwang}
\affiliation{Korea Astronomy and Space Science Institute, 776 Daedeokdae-Ro, Yuseong-Gu, Daejeon 34055, Republic of Korea}




\begin{abstract}
M85 is a peculiar S0 galaxy in Virgo and is a well-known merger remnant.
In this paper, we present the first spectroscopic study of globular clusters (GCs) in M85.
We obtain spectra for 21 GC candidates and the nucleus of M85 using the Gemini Multi-Object Spectrograph on the Gemini North 8.1 m telescope.
From their radial velocities, 20 of the GCs are found to be members of M85.
We find a strong rotation signal of the M85 GC system with a rotation amplitude of 235 \kmsend. 
The rotation axis of the GC system has a position angle of about 161$\arcdeg$, which is 51$\fdg$5 larger than that of the stellar light.
The rotation-corrected radial velocity dispersion of the GC system is estimated to be $\sigma_{\rm r,cor} = $ 160 \kmsend. 
The rotation parameter $\Omega R_{\rm icor}/\sigma_{\rm r,cor}$ of the GC system is derived to be 1.47$^{+1.05}_{-0.48}$, which is one of the largest among known early-type galaxies.
The ages and metallicities of the GCs, which show the same trend as the results based on Lick indices, are derived from full spectrum fitting (ULySS).
About a half of the GCs are an intermediate-age population of which the mean age is $\sim$ 3.7 $\pm$ 1.9 Gyr, having a mean [Fe/H] value of --0.26. The other half are old and metal-poor.
These results suggest that M85 experienced a wet merging event about 4 Gyr ago, forming a significant population of star clusters. The strong rotational feature of the GC system can be explained by an off-center major merging. 
\end{abstract}





\section{I\MakeLowercase{ntroduction}}

M85 \added{(NGC 4382)} is an S0pec galaxy in the northernmost region of the Virgo Cluster and has many interesting and unusual properties.
It has the second highest fine-structure index, $\Sigma = 6.85$, in the list of merger remnant elliptical and S0 galaxies in \citet{ss92}.
The fine-structure index defined by \citet{sch90} is a quantitative parameter that measures how many fine structures exist, which gives information about the time and length of the last merger.
M85 shows isophotes distorted due to its neighbor galaxies \citep{bur79}, a dozen irregular ripples \citep{ss88}, and boxy isophotes within 1$\arcsec$ from the galaxy center\citep{fer06}.
\citet{kor09} found that the surface brightness profile of M85 is different from that of typical S0 galaxies, in the sense that it shows an excess in the outer region at $R\approx1\farcm5$.
M85 shows a positive color gradient in the central region at $R<10''$ (becoming bluer toward the center), which is in contrast to other early-type galaxies \citep{lau05}.
In addition, the nucleus of M85 shows a double structure with a separation of $0\farcs14$ \citep{lau05}.  
\citet{mcd04} revealed, using high-resolution integral-field spectroscopy, that M85 has a counter-rotating kinematically decoupled core within 1$\arcsec$.
All of these features indicate that M85 might have experienced merging events in the recent past.

Several studies have estimated the age of the central stellar light of M85 and concluded that M85 had undergone merger-induced star formation.
\citet{ss92} developed a simple two-burst model for mergers based on the $UBV$ color of galaxies, assuming the merger progenitors and gas-to-star conversion efficiency, and provided a relation between the heuristic merger age and fine-structure index.
From this model, they derived a heuristic merger age for M85 ranging from 3.9 to 7.5 Gyr.
\citet{ffi96} detected a strong H$\beta$ absorption line in the spectrum of the M85 nucleus, which indicates that it may be younger than 3 Gyr.
\citet{tf02} also estimated the age and  metallicity from the H$\beta$ line index and a combination index [MgFe] of the M85 nucleus: an age of 1.6 Gyr and [Fe/H]$=$ 0.44 dex.
These suggest that there was a wet merging event for M85 a few gigayears ago.
{\bf Table \ref{tab:m85.basic}} lists the basic parameters of M85.

\begin{deluxetable}{l c c}
\tabletypesize{\scriptsize}
\tablecaption{Basic Information on M85 \label{tab:m85.basic}}
\tablewidth{\columnwidth}
\tablehead{
\colhead{Parameter} & \colhead{Value} & \colhead{References}
}
\startdata
R.A. (J2000) & 12:25:24.1 & 1 \\
Decl. (J2000) & +18:11:29 & 1 \\
Morphological type & S0$_{1}$(3)pec & 2 \\
$B$-band total magnitude, $B_{T}$ & 10.09 & 2 \\
$V$-band total magnitude, $V_{T}$ & 8.82 & 3 \\
Distance & 17.86 Mpc & 4 \\
Distance modulus, $(m-M)_0$ & 31.26 $\pm$ 0.05 & 4 \\
Foreground reddening, $E(B-V)$ & 0.0239 & 5 \\
Absolute $B$-band total magnitude, $M_B^{T}$ & --21.28 & 2 \\
Absolute $V$-band total magnitude, $M_V^{T}$ & --22.52 & 3 \\
Effective radius from S{\'e}rsic fit, $R_{\rm eff}$ & 128$\farcs$89$^{+10.2}_{-8.8}$ & 3 \\
Image scale & 5.21 kpc arcmin$^{-1}$ & \\
 & 86.8 pc arcsec$^{-1}$ & \\
Radial velocity, \vrad & 729 $\pm$ 2 \kms & 6 \\
\enddata
\vspace{1ex}
\begin{flushleft}
Reference. -- (1) NASA/IPAC Extragalactic Database, (2) \citet{bst85}, (3) \citet{kor09}, (4) \citet{mei07}, (5) \citet{sf11}, (6) \citet{smi00}
\end{flushleft}
\end{deluxetable}

If M85 had experienced a wet merger a few gigayears ago, it is expected that M85 may host numerous intermediate-age globular clusters (GCs) formed during the wet merging process.
There are only two studies of the GCs in M85 in the literature, and they are based purely on photometry.
\citet{pen06} presented the color distribution of the GCs based on the HST/ACS images obtained from the ACS Virgo Cluster Survey \citep{cot04}.
In massive early-type galaxies, GCs often show a bimodality in their color distribution, indicating the presence of two distinct populations: an old metal-poor population and an old metal-rich one.
The color distribution of the GCs in M85, however, is not simply bimodal. \citet{pen06} suggested that M85 is a good candidate that may show a trimodal GC color distribution.
Later, \citet{tra14} investigated the GC population in the northwestern region of M85 with a combination of $gz$ photometry in the ACSVCS and $K_{\rm s}$ photometry obtained using the Near InfraRed Imager and spectrograph on the Gemini North telescope.
They suggested that about 85\% of the observed GCs may be intermediate-age clusters formed about 1.8 Gyr ago.

In this study, we present a spectroscopic survey of the GCs in M85 with the Gemini Multi-Object Spectrograph (GMOS) on the Gemini North telescope.
Spectroscopic surveys of the GCs have advantages over photometric surveys. 
First, spectroscopy is efficient for removing the foreground and background contamination in the photometric GC candidates.
Second, spectral analysis enables us to estimate the ages and metallicities of the GCs independently, breaking the age-metallicity degeneracy involved with photometric studies.
However, to date, there has been no previous spectroscopic study of M85 GCs. 

This paper is organized as follows. 
We briefly describe the spectroscopic target selection, observation, and data reduction in Section 2. 
In Section 3, we derive the radial velocities of targets and identify genuine M85 GCs. 
In Section 4, we investigate the age and metallicity distribution of the M85 GCs and in Section 5, we present the kinematic properties of the M85 GC system.
Primary results are discussed in Section 6.
We summarize the results in Section 7.
We adopted a distance to M85 of 17.9 Mpc \citep{mei07}. One arcmin (one arcsec) corresponds to 5.21 kpc (86.8 pc) at the distance to M85.

\section{O\MakeLowercase{bservation and} d\MakeLowercase{ata} r\MakeLowercase{eduction}}

	\subsection{Target Selection for Spectroscopy}

	\begin{figure}[t]
\epsscale{1}
\includegraphics[trim=20 20 10 20,clip,width=\columnwidth]{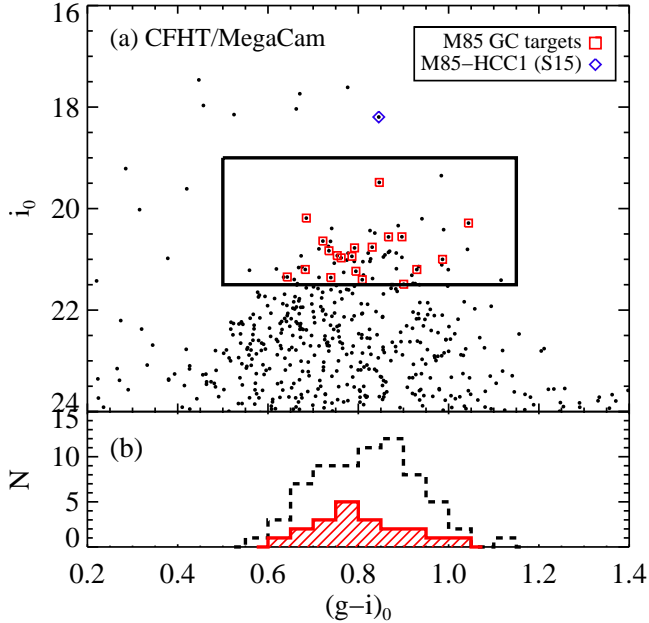}
\caption{(a) $i_0 -(g-i)_0$ CMD of the point sources in the 5$\farcm$5 $\times$ 5$\farcm$5 field around M85..
(b) $(g-i)_0$ color distributions of the observation targets (solid line histogram) and the point sources with the same color and magnitude range as the observation targets (dashed line histogram) in the same field.
The open squares and an open diamond represent the spectroscopic targets, the M85 GC candidates and M85-HCC1 \citep{san15}, respectively.
The bluest of the GC candidates turns out to be a foreground star.
The large box shows the color and magnitude criteria used for the target selection.
\label{fig:target_cmd}}
	\end{figure}

We selected GC candidates from the $ugi$-band images obtained from the CFHT/MegaCam observation (program 14AK06, PI: Myung Gyoon Lee; Ko, Y. et al. 2018, in preparation).
GCs at the distance of the Virgo cluster appear as point sources or slightly extended sources in the CFHT/MegaCam images.
{\bf Figure \ref{fig:target_cmd}(a)} shows the $i_0 - (g-i)_0$ color-magnitude diagram (CMD) for the point sources and the slightly extended sources in the $5\farcm5 \times 5\farcm5$ field centered on M85. 
The $g$ and $i$ magnitudes are based on the CFHT/MegaCam AB system.
The foreground reddening toward M85 is $E(B-V) = 0.024$ \citep{sf11}, and the corresponding extinction values ($A_g = 0.100$  and $A_i = 0.051$) are used in this study.

The dominant population seen in {\bf Figure \ref{fig:target_cmd}(a)} has a color range of $0.5 < (g-i)_0 < 1.15$, and it contains mostly GCs in M85. 
For the GC candidates, we selected the sources with a color range of $0.5 < (g-i)_0 < 1.15$, as done in the case of NGC 474, based on the same magnitude system\citep{lim17}.
Then, we chose the bright GC candidates with $19.0 < i_0 < 21.5$ as the target candidates for spectroscopy, marked by the large box in the figure.
We set the bright limit because all of the sources brighter than $i_0$ = 19.0 mag in the field of view of GMOS are saturated in the CFHT images except for the hypercompact cluster in M85 (M85-HCC1) discovered by \citet{san15}.

Finally, among the bright GC candidates, we selected 21 GC candidates for spectroscopy, while taking into consideration the silt configuration in the Gemini-N/GMOS mask.
\textbf{Figure \ref{fig:target_cmd}(b)} shows the $(g-i)_0$ color distribution of the selected spectroscopic targets, in comparison with that of all GC candidates with the same magnitude and color ranges.  

In addition to our M85 GC candidates, we included two sources as a reference for radial velocity: M85-HCC1, with $g_0=19.134$ mag \citep{san15}, and SDSS J122519.52+181053.7, classified as a star in SDSS DR12 \citep{ala15} with $g_0=17.835$ mag.
We also included the nucleus of M85 for comparison with the M85 GCs.
The basic information for the spectroscopic targets is listed in {\bf Table \ref{tab:obs.target}}.

\begin{deluxetable*}{c c c c c c c c}
\tabletypesize{\scriptsize}
\tablecaption{Photometric Properties and Radial Velocities of the Observation Targets \label{tab:obs.target}}
\tablewidth{\textwidth}
\tablehead{
\colhead{ID} & \colhead{$\alpha$ (J2000)} & \colhead{$\delta$ (J2000)} & \colhead{$g^a$} & \colhead{$(g-i)^a$} & \colhead{$g^b$} & \colhead{$(g-z)^b$} & \colhead{\vrad} \\
\colhead{} & \colhead{(deg)} & \colhead{(deg)} & \colhead{(mag)} & \colhead{(mag)} & \colhead{(mag)} & \colhead{(mag)} & \colhead{(\kmsend)} 
}
\startdata
M85-GC01 & 186.305710 & 18.160259 & 22.087 $\pm$ 0.008 & 1.04 $\pm$ 0.01 & $\cdots$ & $\cdots$ & 1023 $\pm$ 17 \\
M85-GC02 & 186.307587 & 18.206360 & 22.198 $\pm$ 0.008 & 0.79 $\pm$ 0.01 & $\cdots$ & $\cdots$ & 967 $\pm$ 30 \\
M85-GC03 & 186.320374 & 18.160419 & 22.130 $\pm$ 0.008 & 0.84 $\pm$ 0.01 & $\cdots$ & $\cdots$ & 718 $\pm$ 53 \\
M85-GC04 & 186.326767 & 18.193560 & 21.982 $\pm$ 0.007 & 0.73 $\pm$ 0.01 & $\cdots$ & $\cdots$ & 1121 $\pm$ 24 \\
M85-GC05 & 186.329453 & 18.159719 & 21.667 $\pm$ 0.006 & 0.84 $\pm$ 0.01 & $\cdots$ & $\cdots$ & 760 $\pm$ 32 \\
M85-GC06 & 186.333328 & 18.195629 & 21.525 $\pm$ 0.005 & 0.92 $\pm$ 0.01 & 21.475 $\pm$ 0.021 & 1.16 $\pm$ 0.02 & 1027 $\pm$ 17 \\
M85-GC07 & 186.336823 & 18.184280 & 21.833 $\pm$ 0.006 & 0.81 $\pm$ 0.01 & 21.830 $\pm$ 0.021 & 0.97 $\pm$ 0.02 & 555 $\pm$ 15 \\
M85-GC08 & 186.340759 & 18.196980 & 21.430 $\pm$ 0.005 & 1.09 $\pm$ 0.01 & 21.076 $\pm$ 0.027 & 1.19 $\pm$ 0.04 & 659 $\pm$ 16 \\
M85-GC09 & 186.348480 & 18.175610 & 21.827 $\pm$ 0.006 & 0.84 $\pm$ 0.01 & 21.857 $\pm$ 0.016 & 1.05 $\pm$ 0.02 & 915 $\pm$ 23 \\
M85-GC10 & 186.352448 & 18.198999 & 20.431 $\pm$ 0.002 & 0.90 $\pm <$0.01 & 20.330 $\pm$ 0.025 & 1.19 $\pm$ 0.03 & 652 $\pm$ 19 \\
M85-GC11 & 186.354111 & 18.182461 & 21.552 $\pm$ 0.005 & 0.95 $\pm$ 0.01 & 21.583 $\pm$ 0.020 & 1.23 $\pm$ 0.02 & 653 $\pm$ 18 \\
M85-GC12 & 186.356293 & 18.172230 & 21.462 $\pm$ 0.005 & 0.77 $\pm$ 0.01 & 21.509 $\pm$ 0.007 & 0.87 $\pm$ 0.01 & 674 $\pm$ 27 \\
M85-GC13 & 186.360748 & 18.206301 & 21.780 $\pm$ 0.006 & 0.80 $\pm$ 0.01 & 21.836 $\pm$ 0.018 & 1.00 $\pm$ 0.02 & 464 $\pm$ 26 \\
M85-GC14 & 186.365677 & 18.186300 & 21.691 $\pm$ 0.006 & 0.88 $\pm$ 0.01 & 21.695 $\pm$ 0.015 & 1.13 $\pm$ 0.02 & 700 $\pm$ 15 \\
M85-GC15 & 186.368820 & 18.185720 & 20.972 $\pm$ 0.004 & 0.73 $\pm <$0.01 & 20.951 $\pm$ 0.018 & 0.85 $\pm$ 0.02 & 927 $\pm$ 36 \\
M85-GC16 & 186.373337 & 18.193729 & 21.666 $\pm$ 0.006 & 0.78 $\pm$ 0.01 & 21.827 $\pm$ 0.009 & 1.06 $\pm$ 0.03 & 656 $\pm$ 11 \\
M85-GC17 & 186.380371 & 18.205959 & 22.230 $\pm$ 0.009 & 0.98 $\pm$ 0.01 & 22.285 $\pm$ 0.023 & 1.31 $\pm$ 0.03 & 427 $\pm$ 18 \\
M85-GC18 & 186.387253 & 18.186489 & 22.310 $\pm$ 0.009 & 0.86 $\pm$ 0.01 & $\cdots$ & $\cdots$ & 537 $\pm$ 15 \\
M85-GC19 & 186.389374 & 18.206619 & 22.091 $\pm$ 0.008 & 0.69 $\pm$ 0.01 & 22.270 $\pm$ 0.028 & 0.90 $\pm$ 0.03 & 404 $\pm$ 29 \\
M85-GC20 & 186.394760 & 18.170530 & 22.491 $\pm$ 0.010 & 0.95 $\pm$ 0.01 & $\cdots$ & $\cdots$ & 640 $\pm$ 12 \\
Star01 & 186.323792 & 18.172689 & 21.874 $\pm$ 0.007 & 0.61 $\pm$ 0.01 & $\cdots$ & $\cdots$ & 32 $\pm$ 42 \\
M85-HCC1 & 186.345184 & 18.181549 & 19.141 $\pm$ 0.001 & 0.89 $\pm <$0.01 & 19.134 $\pm$ 0.026 & 1.18 $\pm$ 0.04 & 659 $\pm$ 4$^{\rm c}$ \\
M85-Nucleus & 186.350327 & 18.191050 & $\cdots$ & $\cdots$ & $\cdots$ & $\cdots$ & 756 $\pm$ 8 \\
\enddata
%\tablecomments{$^{a}$ Heliocentric radial velocity of M85-HCC1 presented by SDSS.}
%% Text for table notes should follow after the \enddata but before
%% the \end{deluxetable}. Make sure there is at least one \tablenotemark
%% in the table for each \tablenotetext.
\tablenotetext{a}{CFHT/MegaCam AB magnitudes.}
\tablenotetext{b}{HST/ACS AB magnitudes \citep{jor09}.}
\tablenotetext{c}{Heliocentric radial velocity of M85-HCC1 presented by SDSS DR12.}
\end{deluxetable*}

	\subsection{Spectroscopic Observation and Data Reduction}

	\begin{figure}[hbt]
\epsscale{1}
\includegraphics[trim=20 20 20 20,clip,width=\columnwidth]{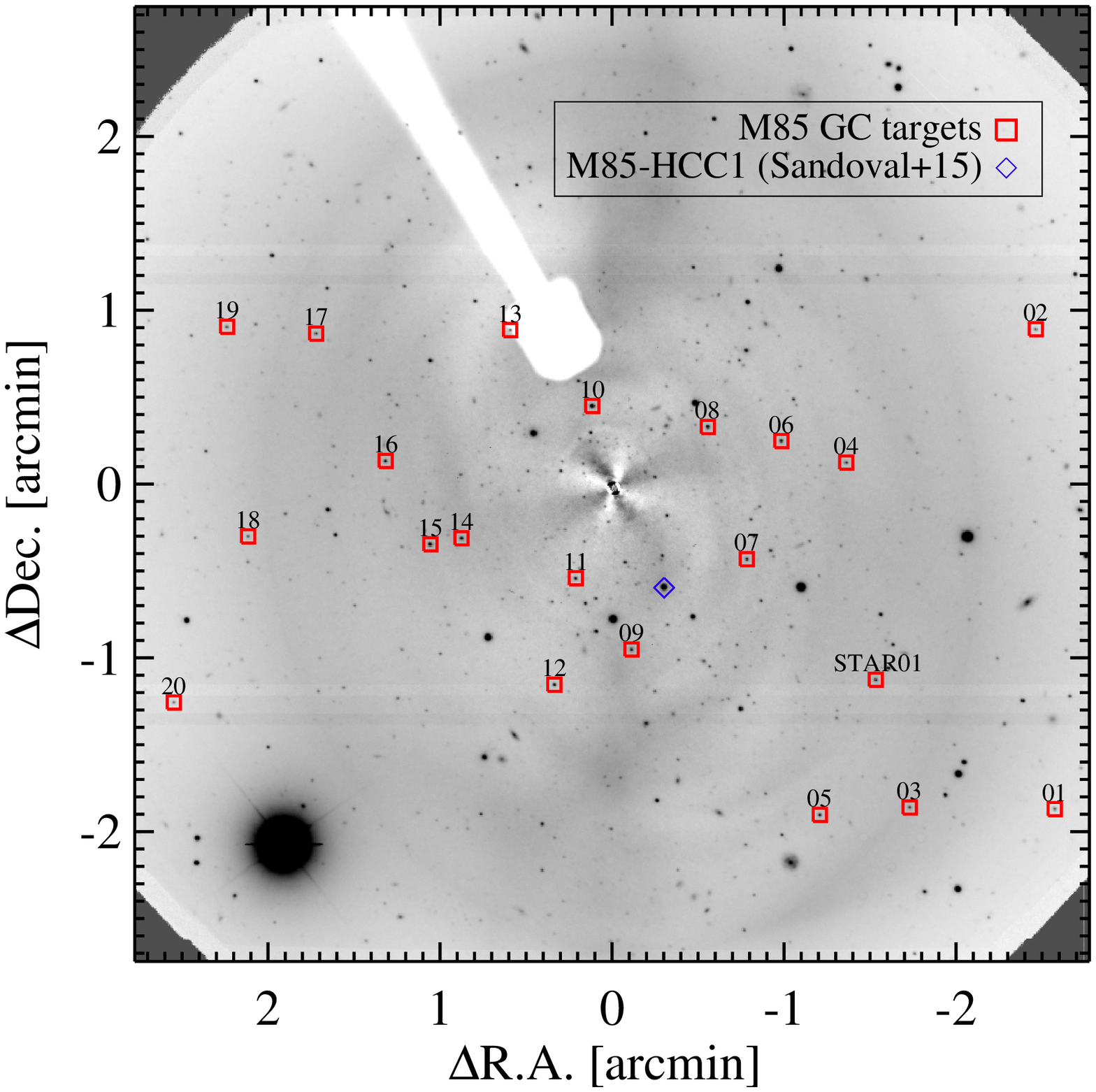}
\caption{Spatial distribution of the spectroscopic targets of GC candidates overlaid
on the GMOS $i$-band image of M85.
The galaxy light of M85 was subtracted from the original image using IRAF/ELLIPSE fitting.
The symbols have the same meaning as in {\bf Figure \ref{fig:target_cmd}}.
The ID numbers for the GCs are marked on the symbols. The source with \vrad = 32 $\pm$ 42 \kms is identified as ``STAR01".
North is up and east to the left.
\label{fig:mask_design}}
	\end{figure}
			
We carried out spectroscopic observations using the Gemini-N/GMOS on 2015 March 22 (program ID: GN-2015A-Q-207, PI: Myung Gyoon Lee).
We designed a mask with a field of view of $5\farcm5 \times 5\farcm5$.
\textbf{Figure \ref{fig:mask_design}} shows the positions of the spectroscopic targets overlaid on the GMOS $i$-band image of M85.
The galaxy light of M85 was subtracted from the original image using IRAF/ELLIPSE fitting.
The spectroscopic targets are located along the E--W direction to avoid slit collision.
We used a B600\_G5307 grating with a dispersion of 0.92 {\rm \AA} pixel$^{-1}$, which covers the wavelength range of about 3800 {\rm \AA} to 6000 {\rm \AA} for most of our targets.
We did a 2 pixel binning in the spectral direction and a 4 pixel binning in the spatial direction for each target in order to improve the signal-to-noise ratio of the observed spectra.
Each target was placed at the central 1$\farcs$0 wide slit.
The exposures were taken in eight sets of 1800 s, and the total integrated exposure time is 4 hr.

We used the Gemini package in IRAF for data reduction.\footnote{IRAF is distributed by the National Optical Astronomy Observatory, which is operated by the Association of Universities for Research in Astronomy (AURA) under a cooperative agreement with the National Science Foundation.}
We used the \textsc{gsflat} task to make a master flat and the \textsc{gsreduce} task for the bias and overscan correction, trimming, and flat-fielding.
The wavelength calibration was done with the \textsc{gswavelength} task using CuAr arc spectra. 
The wavelength solution was applied to the science data using the \textsc{gstransform} task.
We used the \textsc{gemcombine} task to combine eight exposures for each target, and the \textsc{gsskysub} task to subtract the background level.
The spectra were traced and extracted using the \textsc{gsextract} task.
We derived a sensitivity function from the spectrophotometric standard star Hiltner 600, observed during the same night, using \textsc{gsstandard}.
Science spectra were flux-calibrated with \textsc{gscalibrate} using the derived sensitivity function.
The mean signal-to-noise ratio of the final spectra of GC candidates at 4000 -- 5700 $\rm{\AA}$ ranges from 12 to 56.

\section{R\MakeLowercase{adial} v\MakeLowercase{elocity} m\MakeLowercase{easurements and} m\MakeLowercase{embership}}

	\begin{figure}[hbt]
\epsscale{1}
\includegraphics[trim=20 20 10 20,clip,width=\columnwidth]{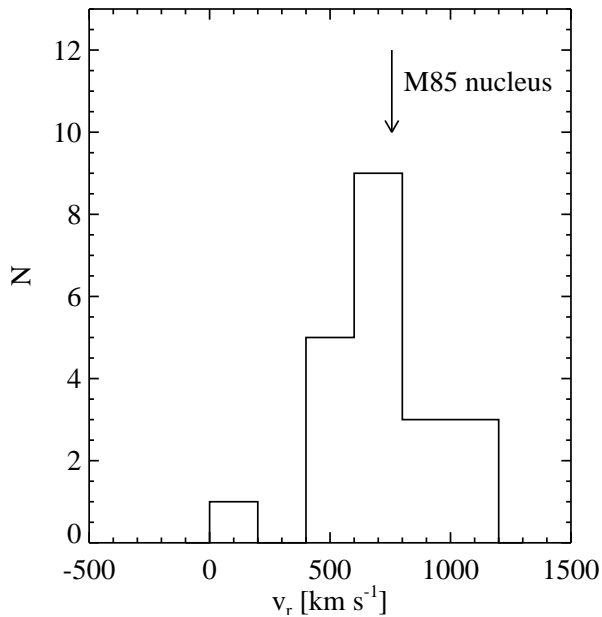}
\caption{Radial velocity distribution of the 21 GC candidates.
Note that the target with \vrad $<$ 100 \kms is considered as a foreground star.
The arrow represents the radial velocity of the M85 nucleus (\vrad = 756 \kmsend).
\label{fig:v_dist1}}	
	\end{figure}
	
	\begin{figure}
\epsscale{1}
\includegraphics[trim= 20 0 10 0,clip,width=\columnwidth]{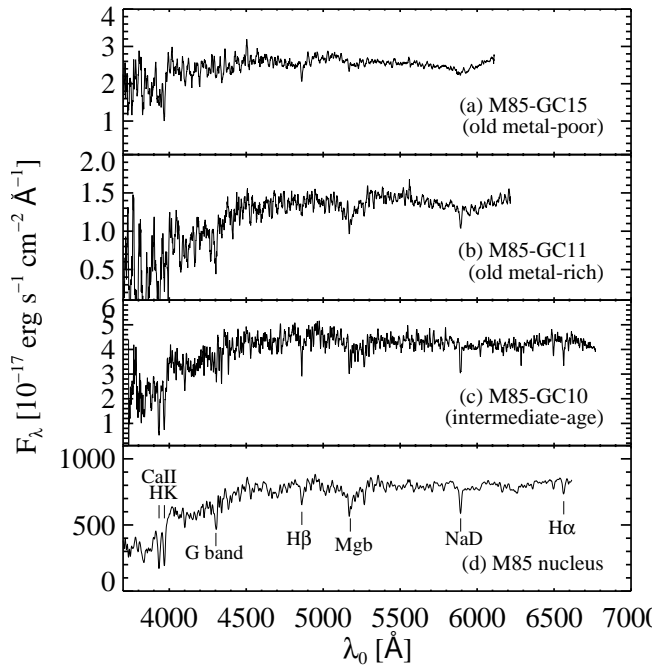}
\caption{Sample spectra of 
(a) an old metal-poor GC (ID: M85-GC15) with age of 13 Gyr and [Fe/H] of --1.42,
(b) an old metal-rich GC (ID: M85-GC11) with age of 15 Gyr and [Fe/H] of --0.50, 
(c) an intermediate-age GC (ID: M85-GC10) with age of 2 Gyr and [Fe/H] of 0.20, and (d) the M85 nucleus. 
All flux-calibrated spectra are plotted in the rest frame and smoothed using a boxcar filter with a size of 10 \rm{\AA}.
\label{fig:sample_spec}}
	\end{figure}

	\begin{figure*}
\epsscale{1}
\includegraphics[width=\textwidth]{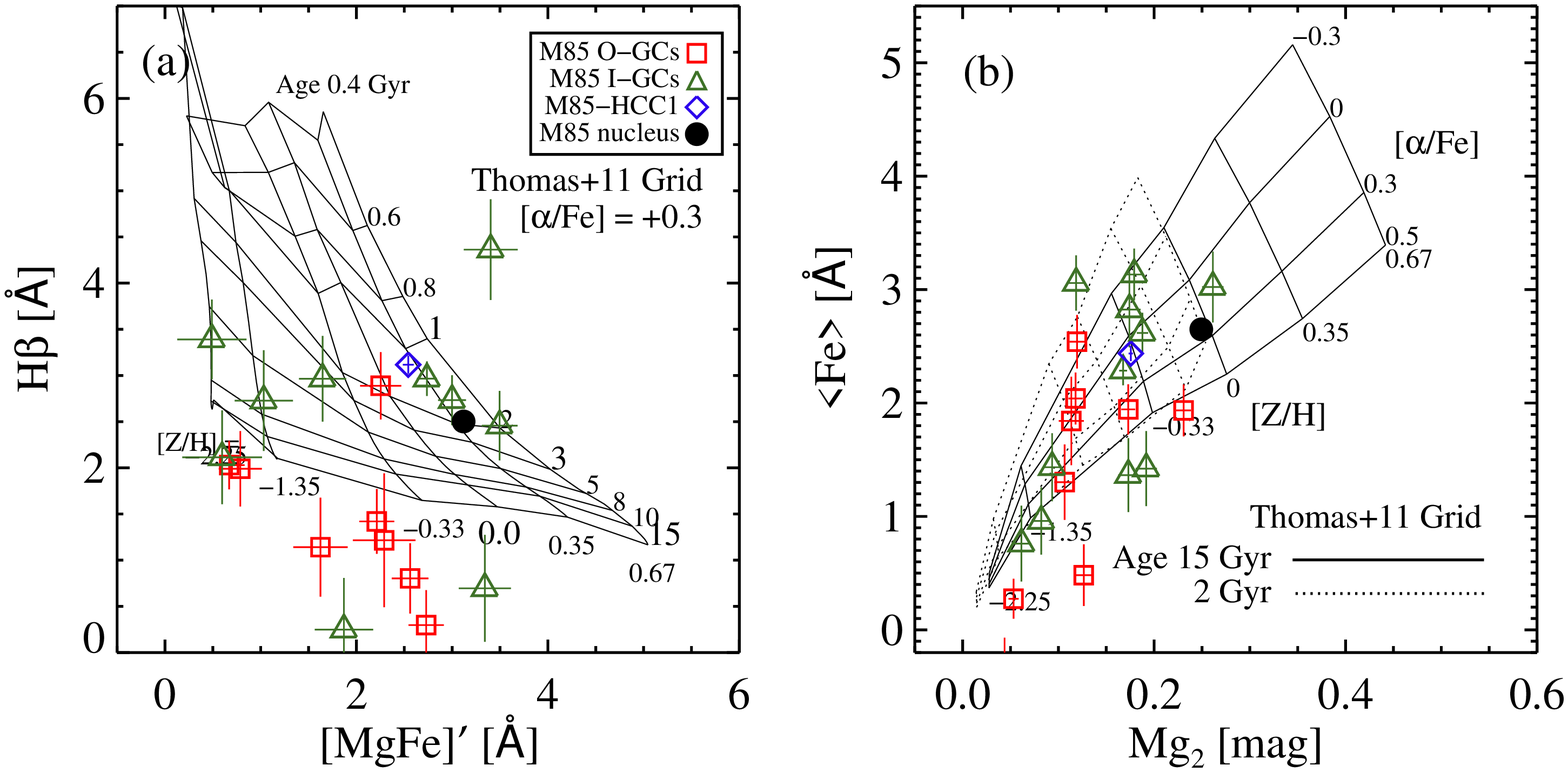}
\caption{(a) Lick line indices for the H$\beta$ line vs. [MgFe]$\arcmin$ for the M85 GCs (red squares and green triangles), M85-HCC1 (blue diamond), and the M85 nucleus (black filled circle) measured with EZ\_Ages \citep{gs08} in this study. \added{The red squares and green triangles represent the old and intermediate-age GCs classified based on the ages from the full spectrum fitting method, respectively.}
The error bars represent the 1$\sigma$ uncertainties of the Lick line indices.
The grids present the SSP models for a given [$\alpha$/Fe] of +0.3 and for various values of [Z/H] (--2.25, --1.35, --0.33, 0, +0.33, and +0.67) and ages (0.4, 0.6, 0.8, 1, 2, 3, 5, 8, 10, and 15 Gyr) of \citet{tmj11}.
(b) $<$Fe$>$ vs. Mg$_2$. The symbols are the same as in (a). The solid and dotted line grids are for a given age of 15 Gyr and 2 Gyr, respectively.
\label{fig:lick_diagram}}
	\end{figure*}
		
We used the Fourier cross-correlation task, \textsc{fxcor}, in the IRAF \textsc{rv} package \citep{td79}, to estimate the heliocentric radial velocities of the spectroscopic targets.
SDSS J122519.52+181053.7 and M85-HCC1 are used as spectral templates for the radial velocity measurements.
The radial velocities of the SDSS star and M85-HCC1 are 35 $\pm$ 4 \kms and 659 $\pm$ 4 \kmsend, respectively, according to the SDSS DR12.
To apply the cross-correlation method, we used the prominent absorption features of the spectroscopic targets over the wavelength range of 4840 -- 5500 \AA.
For each target, we adopted the error-weighted mean value of the radial velocities determined by two spectral templates. 

We measured the radial velocity of the M85 nucleus to be 756 $\pm$ 8 \kms, similar to the previous measurements, \vrad = 729 $\pm$ 2 \kms from \citet{smi00} and \vrad = 760 \kms from \citet{gav04}.
The radial velocity uncertainties range from 8 to 53 \kmsend, with a mean value of 23 $\pm$ 11 \kmsend.

\textbf{Figure \ref{fig:v_dist1}} shows the heliocentric radial velocity distribution of 21 GC candidates.
The radial velocity distribution of most GC candidates shows a peak at 700 \kmsend, which is similar to that of the M85 nucleus. Only one GC candidate shows a large deviation, at \vrad = 32 \kmsend.  
We considered 20 GC candidates with \vrad = 404 -- 1121 \kms to be the GCs bound to M85.
The source with \vrad = 32 \kms is considered to be a foreground star. 

\textbf{Figure \ref{fig:sample_spec}} shows the spectra of three GC examples (an old metal-poor GC, M85-GC15; an old metal-rich GC, M85-GC11; and an intermediate-age GC, M85-GC10) confirmed in this study, in comparison with the spectrum of the M85 nucleus.
The continuum of the old metal-poor GC spectrum is flatter than that of the old metal-rich GC and intermediate-age GC spectra.
The old metal-rich GC spectrum shows a stronger Mg$b$ absorption line, when compared with the old metal-poor GC spectrum.
The intermediate-age GC spectrum shows stronger H$\beta$ and Mg$b$ absorption lines, resembling the spectrum of the M85 nucleus.

\section{A\MakeLowercase{ge}, [Z/H], and [$\alpha$/F\MakeLowercase{e}] of M85 GC\MakeLowercase{s}}

	\subsection{Lick Indices}

\citet{bur84} measured the strengths of 11 prominent absorption lines in the spectra of 17 Galactic GCs with a resolution of about 9 \rm{\AA} and introduced them as the Lick index system.
Later, \citet{wor94} and \citet{wo97} added 10 and 4 more absorption lines to the Lick index system, respectively. 
\citet{tra98} refined the system with a definition of these 25 absorption lines.
The Lick line indices are widely used to determine the ages and metallicities of old stellar systems by comparing them with those expected from theoretical models.

We measured the Lick indices from the spectra of the 20 GCs, M85-HCC1, and the M85 nucleus using the EZ\_Ages package \citep{gs08} based on the stellar population model of \citet{sch07}.
Since we did not observe the Lick standard stars during our observing run, we could not calibrate our values of the Lick indices to the Lick standard system.
\citet{pie06a,pie06b} also used noncalibrated Lick indices to determine the ages and metallicities of GCs in two elliptical galaxies, NGC 3379 and M60. They noticed that some indices show systematic differences from those expected from simple stellar population (SSP) models. We also caution about this point.
The Lick line indices and errors of all observation targets are listed in {\bf Tables \ref{tab:lick.index} and \ref{tab:lick.index.err}}, respectively.

We used two independent methods based on these Lick line indices to derive the ages and metallicities of M85 GCs, M85-HCC1, and the M85 nucleus: a Lick index grid method and a $\chi^2$ minimization method.
We used the SSP models of \citet{tmj11}.
The ages of the SSP models of \citet{tmj11} range from 0.1 to 15 Gyr, the metallicities [Z/H] from --2.25 to +0.67, and the $\alpha$-element abundances [$\alpha$/Fe] from --0.3 to +0.5.

	\subsubsection{Lick Index Grid Method}
	
The Lick index grid method uses age- and metallicity-sensitive absorption lines such as Balmer lines (H$\beta$, H$\gamma$, and H$\delta$), Fe5270, Fe5335, and Mg$b$ lines \citep{puz05}.
{\bf Figure \ref{fig:lick_diagram}} shows the diagnostic grids with the Lick indices of the M85 GCs, M85-HCC1, and the M85 nucleus: (a) H$\beta$ vs. [MgFe]$'$ and (b) $<$Fe$>$ vs. Mg$_2$.
We selected the H$\beta$ index for the age indicator rather than the H$\gamma$ and H$\delta$ indices because of its higher signal-to-noise ratio.
The [MgFe]$'$ index is defined as [MgFe]$'$ = ${\rm \sqrt{Mgb \times (0.72 \times Fe5270 + 0.28 \times Fe5335)}}$, which is a metallicity-sensitive composite index and little affected by [$\alpha$/Fe].
The $<$Fe$>$ index is a metallicity indicator defined as $<$Fe$>$ = (Fe5270 + Fe5335)/2, while the Mg$_2$ index is sensitive to $\alpha$-element abundances.

In {\bf Figure \ref{fig:lick_diagram}(a)}, the sample of the M85 GCs is divided into two groups: an intermediate-age group with ages of $\sim$ 1-2 Gyr and an old group with ages older than 10 Gyr. The old GCs are all located below the 15 Gyr limit of the model in the figure.
Most GCs older than 5 Gyr have metallicities lower than the solar metallicity ([Z/H] = 0).
M85-HCC1 and the M85 nucleus have ages of a few gigayears and supersolar metallicities.
A few M85 GCs show ages and metallicities similar to those of the M85 nucleus.
{\bf Figure \ref{fig:lick_diagram}(b)} shows that the [$\alpha$/Fe] values of the GCs range from --0.3 to +0.3.
Most of the GCs in M85 are on the grids.
We determined the ages, metallicities, and [$\alpha$/Fe] of all GCs, M85-HCC1, and the M85 nucleus through iterations between two diagnostic grids: the H$\beta$ versus [MgFe]$'$ grid and the Mg$_2$ versus $<$Fe$>$ grid.
We followed the iteration technique described in \citet{puz05} and \citet{par12}.
The [Z/H] can be measured from each grid, so we present the [Z/H] values from both grids.
Note that there are some GCs older than 15 Gyr that fall outside the diagnostic grid.
In this case, we adopted the parameters of the nearest point of the model grids in the direction of their error vector.	

	\subsubsection{$\chi^2$ Minimization Method}
	
\citet{pro04} suggested a $\chi^2$ minimization technique based on the residuals between the observed Lick indices and SSP model predictions.
This method uses as many Lick indices as possible so that it does not depend significantly on any specific lines.
Among the 25 Lick indices, we excluded several lines in each spectrum for the fitting.
First, we excluded the CN$_1$, CN$_2$, Ca4227, and NaD indices, following \citet{pro04}.
The nitrogen abundance anomaly is a well-known problem for Galactic GCs.
The CN indices and Ca4227 index are sensitive to nitrogen abundances, so they do not fit well with the typical SSP models.
The NaD index is rejected because it is severely affected by interstellar absorption.
Second, we only used the absorption lines with signal-to-noise ratios higher than 10.
Third, we excluded some indices after $\sim 2\sigma$ clipping of their $\chi$ values iteratively. 
Finally, we used 8--20 Lick indices of each GC for the fitting.
	
	\begin{figure}
\epsscale{1}
\includegraphics[width=\columnwidth]{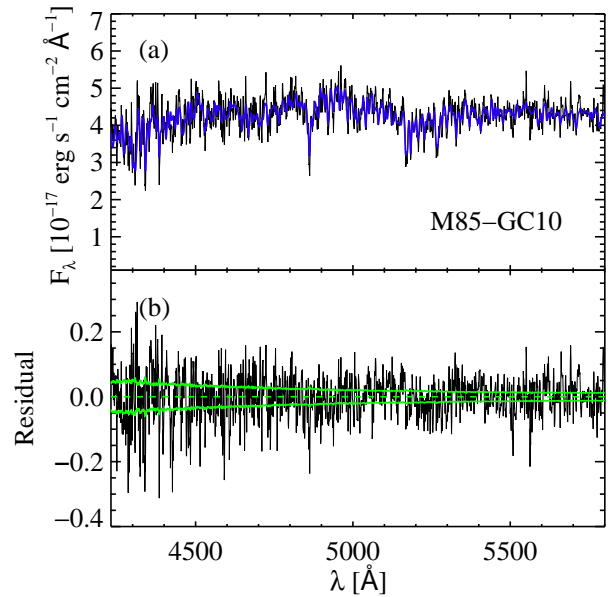}
\caption{(a) Full spectrum fitting result for M85-GC10. 
The black and blue lines indicate the spectra of M85-GC10 and the best fit from full spectrum fitting results of ULySS, respectively.
(b) Residuals from the best fit (black line).
The green solid lines represent the $\pm1\sigma$ error residuals corresponding to the input signal-to-noise ratio.
\label{fig:fitres}}
	\end{figure}

\begin{turnpage}
\begin{deluxetable*}{l c c c c c c c c c c c c c c c c c c c c c c c c c}
\tabletypesize{\tiny}
\setlength{\tabcolsep}{0.02in}
\tablecaption{Lick Line Indices of the M85 GCs, M85-HCC1, and the M85 Nucleus \label{tab:lick.index}}
\tablewidth{0pt}
\tablehead{
\colhead{ID} & \colhead{CN$_1$} & \colhead{CN$_2$} & \colhead{Ca4227} & \colhead{G4300} & \colhead{Fe4383} & \colhead{Ca4455} & \colhead{Fe4531} & \colhead{Fe4668} & \colhead{H$\beta$} & \colhead{Fe5015} & \colhead{Mg$_1$} & \colhead{Mg$_2$} & \colhead{Mg$b$} & \colhead{Fe5270} & \colhead{Fe5335} & \colhead{Fe5406} & \colhead{Fe5709} & \colhead{Fe5782} & \colhead{NaD} & \colhead{TiO$_1$} & \colhead{TiO$_2$} & \colhead{H$\delta_A$} & \colhead{H$\gamma_A$} & \colhead{H$\delta_F$} & \colhead{H$\gamma_F$} \\
\colhead{} & \colhead{(mag)} & \colhead{(mag)} & \colhead{(\AA)} & \colhead{(\AA)} & \colhead{(\AA)} & \colhead{(\AA)} & \colhead{(\AA)} & \colhead{(\AA)} & \colhead{(\AA)} & \colhead{(\AA)} & \colhead{(mag)} & \colhead{(mag)} & \colhead{(\AA)} & \colhead{(\AA)} & \colhead{(\AA)} & \colhead{(\AA)} & \colhead{(\AA)} & \colhead{(\AA)} & \colhead{(\AA)} & \colhead{(mag)} & \colhead{(mag)} & \colhead{(\AA)} & \colhead{(\AA)} & \colhead{(\AA)} & \colhead{(\AA)}
}
\startdata
M85-GC01 & $ -0.149$ & $ -0.182$ & $ -3.312$ & $  4.752$ & $  3.471$ & $  2.670$ & $ -0.179$ & $ 10.335$ & $  0.695$ & $  4.015$ & $  0.149$ & $  0.261$ & $  3.255$ & $  3.950$ & $  2.096$ & $  2.831$ & $  0.870$ & $ -0.378$ & $  3.133$ & $\cdots$ & $\cdots$ & $  2.415$ & $ -9.828$ & $  7.512$ & $ -1.894$ \\
M85-GC02 & $  0.113$ & $  0.159$ & $ -0.048$ & $  4.686$ & $  3.881$ & $  4.768$ & $  6.288$ & $ -2.627$ & $  1.141$ & $  3.619$ & $ -0.027$ & $  0.107$ & $  2.194$ & $  1.076$ & $  1.529$ & $  0.941$ & $  0.192$ & $ -0.080$ & $ -0.510$ & $  0.053$ & $ -0.014$ & $ -1.683$ & $  0.274$ & $  2.471$ & $  0.766$ \\
M85-GC03 & $  0.005$ & $  0.171$ & $ -7.530$ & $  9.217$ & $  7.369$ & $  6.379$ & $ -3.273$ & $  4.157$ & $  2.727$ & $  1.206$ & $  0.097$ & $  0.192$ & $  0.922$ & $  0.815$ & $  2.028$ & $  0.693$ & $  0.624$ & $  0.518$ & $  2.683$ & $\cdots$ & $\cdots$ & $ -0.607$ & $ -5.127$ & $ -2.534$ & $  2.505$ \\
M85-GC04 & $ -0.182$ & $ -0.157$ & $ -2.725$ & $  2.068$ & $  5.928$ & $  0.190$ & $  2.020$ & $ -7.188$ & $  2.965$ & $  0.587$ & $ -0.032$ & $  0.093$ & $  1.958$ & $  1.323$ & $  1.546$ & $  1.323$ & $  0.668$ & $ -0.723$ & $ -0.946$ & $  0.042$ & $ -0.011$ & $  4.474$ & $ -0.413$ & $  0.693$ & $  1.383$ \\
M85-GC05 & $ -0.174$ & $ -0.147$ & $ -1.346$ & $  5.066$ & $  1.875$ & $ -0.828$ & $  0.820$ & $  3.521$ & $  1.407$ & $  8.384$ & $  0.093$ & $  0.119$ & $ -0.310$ & $  3.365$ & $  2.752$ & $  2.135$ & $  1.012$ & $  0.317$ & $  3.392$ & $\cdots$ & $\cdots$ & $  4.396$ & $ -7.127$ & $  4.040$ & $  0.061$ \\
M85-GC06 & $  0.011$ & $  0.045$ & $  1.220$ & $ -2.780$ & $  6.401$ & $ -0.911$ & $  3.525$ & $  4.286$ & $  1.419$ & $  2.448$ & $  0.018$ & $  0.173$ & $  2.564$ & $  1.867$ & $  2.021$ & $  1.328$ & $  0.146$ & $  0.186$ & $  0.163$ & $  0.056$ & $ -0.037$ & $ -5.655$ & $ -2.775$ & $ -0.972$ & $ -1.117$ \\
M85-GC07 & $ -0.108$ & $ -0.026$ & $  1.982$ & $ -1.510$ & $  2.896$ & $  1.476$ & $  5.552$ & $ -1.293$ & $  1.990$ & $ -0.942$ & $  0.031$ & $  0.126$ & $  0.963$ & $  0.848$ & $  0.116$ & $  1.180$ & $  1.328$ & $  0.129$ & $  2.314$ & $  0.058$ & $ -0.047$ & $  9.068$ & $  2.020$ & $  5.998$ & $  0.049$ \\
M85-GC08 & $  0.055$ & $  0.096$ & $  0.269$ & $  3.784$ & $  3.767$ & $  0.282$ & $  3.481$ & $  9.609$ & $  2.732$ & $  4.731$ & $  0.044$ & $  0.188$ & $  3.249$ & $  2.961$ & $  2.275$ & $  1.920$ & $  1.097$ & $  0.496$ & $  3.321$ & $  0.094$ & $  0.006$ & $  1.946$ & $ -2.259$ & $  2.986$ & $  1.832$ \\
M85-GC09 & $ -0.217$ & $  0.112$ & $  0.173$ & $  0.916$ & $  4.541$ & $ -0.081$ & $ -2.336$ & $ -1.322$ & $  3.390$ & $ -0.092$ & $  0.047$ & $  0.082$ & $  0.275$ & $  0.762$ & $  1.159$ & $  1.186$ & $  1.457$ & $  0.138$ & $  2.075$ & $  0.094$ & $\cdots$ & $ 14.329$ & $ -2.478$ & $  5.196$ & $  2.019$ \\
M85-GC10 & $  0.055$ & $  0.070$ & $  0.049$ & $  1.199$ & $  4.029$ & $  1.111$ & $  2.650$ & $  4.625$ & $  2.965$ & $  4.717$ & $  0.016$ & $  0.168$ & $  2.980$ & $  2.801$ & $  1.770$ & $  0.801$ & $  0.215$ & $  0.498$ & $  3.082$ & $  0.055$ & $  0.028$ & $  0.660$ & $ -1.502$ & $  1.487$ & $  1.660$ \\
M85-GC11 & $ -0.064$ & $ -0.012$ & $ -0.860$ & $  5.207$ & $  1.427$ & $  1.295$ & $  2.357$ & $  1.495$ & $  0.803$ & $  1.795$ & $  0.083$ & $  0.231$ & $  3.067$ & $  2.397$ & $  1.473$ & $  1.160$ & $  0.640$ & $  0.693$ & $  2.416$ & $  0.074$ & $  0.225$ & $  6.045$ & $ -4.265$ & $ -0.867$ & $  0.066$ \\
M85-GC12 & $ -0.106$ & $ -0.066$ & $ -2.580$ & $ -1.425$ & $  0.630$ & $ -0.564$ & $ -0.838$ & $  4.041$ & $  3.293$ & $  0.077$ & $  0.035$ & $  0.044$ & $  0.181$ & $ -1.202$ & $  0.605$ & $ -0.480$ & $  0.186$ & $ -0.431$ & $  1.654$ & $  0.014$ & $\cdots$ & $  0.482$ & $  0.993$ & $ -2.155$ & $  0.217$ \\
M85-GC13 & $ -0.097$ & $ -0.116$ & $  0.974$ & $  6.720$ & $  8.688$ & $  0.812$ & $  0.130$ & $ -4.409$ & $  0.297$ & $  2.183$ & $  0.007$ & $  0.118$ & $  3.186$ & $  2.717$ & $  1.364$ & $  0.108$ & $  1.013$ & $  0.314$ & $ -1.136$ & $ -0.008$ & $ -0.032$ & $  0.927$ & $ -6.750$ & $ -0.489$ & $  1.235$ \\
M85-GC14 & $ -0.124$ & $ -0.114$ & $  2.476$ & $ -8.870$ & $  7.809$ & $  3.071$ & $  3.995$ & $  5.194$ & $  2.889$ & $  2.716$ & $ -0.025$ & $  0.119$ & $  1.874$ & $  2.926$ & $  2.154$ & $  0.740$ & $  0.829$ & $  0.479$ & $  1.843$ & $  0.076$ & $  0.037$ & $  7.941$ & $  3.289$ & $  3.565$ & $  1.848$ \\
M85-GC15 & $  0.026$ & $  0.071$ & $ -0.712$ & $  1.559$ & $  4.315$ & $  1.483$ & $  3.651$ & $  1.008$ & $  2.030$ & $  1.693$ & $ -0.027$ & $  0.053$ & $  1.063$ & $  0.613$ & $ -0.065$ & $  0.392$ & $  0.203$ & $ -0.227$ & $  0.324$ & $  0.044$ & $ -0.044$ & $ -0.336$ & $ -0.117$ & $  1.893$ & $  1.808$ \\
M85-GC16 & $ -0.022$ & $ -0.017$ & $  2.311$ & $  3.773$ & $  8.241$ & $  2.649$ & $ -0.390$ & $  4.685$ & $  2.457$ & $  4.869$ & $  0.049$ & $  0.179$ & $  4.069$ & $  2.840$ & $  3.426$ & $  0.856$ & $  1.264$ & $  1.354$ & $  3.029$ & $  0.042$ & $ -0.004$ & $  6.423$ & $ -1.889$ & $  3.883$ & $  2.599$ \\
M85-GC17 & $ -0.084$ & $  0.065$ & $ -0.649$ & $ -1.010$ & $ 13.273$ & $  1.128$ & $ -0.859$ & $  0.594$ & $  0.248$ & $  7.803$ & $  0.015$ & $  0.173$ & $  2.972$ & $  0.936$ & $  1.794$ & $  2.132$ & $  1.006$ & $  1.165$ & $  2.627$ & $  0.033$ & $  0.031$ & $  7.567$ & $ -5.257$ & $  4.936$ & $  0.503$ \\
M85-GC18 & $  0.212$ & $ -0.032$ & $ -2.163$ & $  7.700$ & $-12.578$ & $ -0.927$ & $  7.676$ & $ -5.943$ & $  4.363$ & $  5.298$ & $  0.027$ & $  0.174$ & $  4.316$ & $  2.504$ & $  3.144$ & $  0.828$ & $  0.788$ & $  0.641$ & $  0.131$ & $  0.051$ & $  0.029$ & $-10.287$ & $ -3.909$ & $ -3.408$ & $ -1.665$ \\
M85-GC19 & $ -0.080$ & $ -0.019$ & $  1.108$ & $  0.581$ & $ -6.532$ & $ -1.105$ & $  0.473$ & $ -0.478$ & $  2.114$ & $  5.165$ & $ -0.019$ & $  0.061$ & $  1.346$ & $ -0.365$ & $  1.885$ & $ -0.047$ & $ -0.030$ & $  0.269$ & $ -0.850$ & $  0.050$ & $ -0.052$ & $  9.094$ & $ -5.530$ & $  4.158$ & $ -6.059$ \\
M85-GC20 & $  0.160$ & $  0.056$ & $  1.938$ & $  1.920$ & $  2.661$ & $  0.479$ & $  0.103$ & $ -0.820$ & $  1.215$ & $  7.026$ & $  0.035$ & $  0.113$ & $  2.739$ & $  2.006$ & $  1.677$ & $  1.388$ & $  0.791$ & $ -0.614$ & $  3.548$ & $  0.067$ & $\cdots$ & $-13.684$ & $  5.390$ & $ -9.088$ & $  6.152$ \\
M85-HCC1 & $  0.006$ & $  0.039$ & $  0.792$ & $  2.637$ & $  2.869$ & $  0.967$ & $  3.102$ & $  5.823$ & $  3.117$ & $  4.751$ & $  0.041$ & $  0.176$ & $  2.610$ & $  2.523$ & $  2.351$ & $  1.507$ & $  0.820$ & $  0.703$ & $  3.522$ & $  0.060$ & $  0.056$ & $  1.577$ & $ -1.114$ & $  1.855$ & $  1.121$ \\
M85-Nucleus & $  0.053$ & $  0.081$ & $  0.856$ & $  4.423$ & $  3.930$ & $  1.064$ & $  3.383$ & $  7.175$ & $  2.502$ & $  5.349$ & $  0.084$ & $  0.249$ & $  3.563$ & $  2.837$ & $  2.461$ & $  1.595$ & $  0.899$ & $  0.747$ & $  3.775$ & $  0.061$ & $  0.064$ & $  0.244$ & $ -4.117$ & $  1.198$ & $ -0.195$ \\
\enddata
%\tablecomments{$^{a}$ The number of GC candidates assigned to fiber.}
%% Text for table notes should follow after the \enddata but before
%% the \end{deluxetable}. Make sure there is at least one \tablenotemark
%% in the table for each \tablenotetext.
%\tablenotetext{a}{CFHT/MegaCam AB magnitudes.}
\end{deluxetable*}
\end{turnpage}

\begin{turnpage}
\begin{deluxetable*}{l c c c c c c c c c c c c c c c c c c c c c c c c c}
\tabletypesize{\tiny}
\setlength{\tabcolsep}{0.025in}
\tablecaption{Lick Line Index Errors of the M85 GCs, M85-HCC1, and the M85 Nucleus \label{tab:lick.index.err}}
\tablewidth{0pt}
\tablehead{
\colhead{ID} & \colhead{CN$_1$} & \colhead{CN$_2$} & \colhead{Ca4227} & \colhead{G4300} & \colhead{Fe4383} & \colhead{Ca4455} & \colhead{Fe4531} & \colhead{Fe4668} & \colhead{H$\beta$} & \colhead{Fe5015} & \colhead{Mg$_1$} & \colhead{Mg$_2$} & \colhead{Mg$b$} & \colhead{Fe5270} & \colhead{Fe5335} & \colhead{Fe5406} & \colhead{Fe5709} & \colhead{Fe5782} & \colhead{NaD} & \colhead{TiO$_1$} & \colhead{TiO$_2$} & \colhead{H$\delta_A$} & \colhead{H$\gamma_A$} & \colhead{H$\delta_F$} & \colhead{H$\gamma_F$} \\
\colhead{} & \colhead{(mag)} & \colhead{(mag)} & \colhead{(\AA)} & \colhead{(\AA)} & \colhead{(\AA)} & \colhead{(\AA)} & \colhead{(\AA)} & \colhead{(\AA)} & \colhead{(\AA)} & \colhead{(\AA)} & \colhead{(mag)} & \colhead{(mag)} & \colhead{(\AA)} & \colhead{(\AA)} & \colhead{(\AA)} & \colhead{(\AA)} & \colhead{(\AA)} & \colhead{(\AA)} & \colhead{(\AA)} & \colhead{(mag)} & \colhead{(mag)} & \colhead{(\AA)} & \colhead{(\AA)} & \colhead{(\AA)} & \colhead{(\AA)}
}
\startdata
M85-GC01 & $  0.049$ & $  0.064$ & $  1.356$ & $  1.427$ & $  1.757$ & $  0.842$ & $  1.368$ & $  1.537$ & $  0.579$ & $  1.027$ & $  0.010$ & $  0.011$ & $  0.428$ & $  0.420$ & $  0.462$ & $  0.319$ & $  0.231$ & $  0.229$ & $  0.251$ & $\cdots$ & $\cdots$ & $  1.775$ & $  1.574$ & $  0.972$ & $  0.916$ \\
M85-GC02 & $  0.044$ & $  0.053$ & $  0.794$ & $  1.198$ & $  1.826$ & $  0.760$ & $  1.292$ & $  1.793$ & $  0.536$ & $  0.937$ & $  0.010$ & $  0.011$ & $  0.423$ & $  0.448$ & $  0.493$ & $  0.348$ & $  0.246$ & $  0.236$ & $  0.309$ & $  0.006$ & $  0.005$ & $  1.733$ & $  1.330$ & $  1.101$ & $  0.843$ \\
M85-GC03 & $  0.047$ & $  0.055$ & $  1.459$ & $  1.214$ & $  1.635$ & $  0.831$ & $  1.639$ & $  1.704$ & $  0.545$ & $  1.067$ & $  0.010$ & $  0.011$ & $  0.464$ & $  0.452$ & $  0.484$ & $  0.363$ & $  0.252$ & $  0.240$ & $  0.277$ & $\cdots$ & $\cdots$ & $  1.632$ & $  1.595$ & $  1.301$ & $  0.877$ \\
M85-GC04 & $  0.032$ & $  0.040$ & $  0.769$ & $  1.041$ & $  1.315$ & $  0.664$ & $  1.088$ & $  1.585$ & $  0.464$ & $  0.892$ & $  0.009$ & $  0.009$ & $  0.375$ & $  0.409$ & $  0.437$ & $  0.310$ & $  0.214$ & $  0.221$ & $  0.276$ & $  0.006$ & $  0.005$ & $  1.040$ & $  1.028$ & $  0.835$ & $  0.666$ \\
M85-GC05 & $  0.028$ & $  0.036$ & $  0.573$ & $  0.808$ & $  1.200$ & $  0.680$ & $  0.951$ & $  1.238$ & $  0.433$ & $  0.714$ & $  0.008$ & $  0.008$ & $  0.360$ & $  0.328$ & $  0.361$ & $  0.261$ & $  0.195$ & $  0.186$ & $  0.208$ & $\cdots$ & $\cdots$ & $  0.991$ & $  0.925$ & $  0.708$ & $  0.522$ \\
M85-GC06 & $  0.025$ & $  0.031$ & $  0.448$ & $  0.893$ & $  1.066$ & $  0.626$ & $  0.788$ & $  1.005$ & $  0.351$ & $  0.647$ & $  0.006$ & $  0.007$ & $  0.283$ & $  0.299$ & $  0.324$ & $  0.232$ & $  0.171$ & $  0.162$ & $  0.211$ & $  0.004$ & $  0.004$ & $  1.105$ & $  0.793$ & $  0.712$ & $  0.496$ \\
M85-GC07 & $  0.031$ & $  0.038$ & $  0.521$ & $  1.105$ & $  1.299$ & $  0.683$ & $  0.866$ & $  1.329$ & $  0.408$ & $  0.810$ & $  0.008$ & $  0.008$ & $  0.349$ & $  0.363$ & $  0.404$ & $  0.280$ & $  0.194$ & $  0.195$ & $  0.236$ & $  0.005$ & $  0.004$ & $  0.873$ & $  0.913$ & $  0.630$ & $  0.628$ \\
M85-GC08 & $  0.021$ & $  0.026$ & $  0.366$ & $  0.625$ & $  0.834$ & $  0.461$ & $  0.603$ & $  0.745$ & $  0.269$ & $  0.492$ & $  0.005$ & $  0.006$ & $  0.224$ & $  0.241$ & $  0.262$ & $  0.189$ & $  0.137$ & $  0.133$ & $  0.158$ & $  0.004$ & $  0.003$ & $  0.707$ & $  0.645$ & $  0.474$ & $  0.378$ \\
M85-GC09 & $  0.050$ & $  0.056$ & $  0.661$ & $  1.200$ & $  1.358$ & $  0.800$ & $  1.126$ & $  1.478$ & $  0.432$ & $  0.937$ & $  0.009$ & $  0.009$ & $  0.393$ & $  0.396$ & $  0.442$ & $  0.305$ & $  0.213$ & $  0.215$ & $  0.253$ & $  0.006$ & $\cdots$ & $  1.458$ & $  1.090$ & $  1.493$ & $  0.596$ \\
M85-GC10 & $  0.013$ & $  0.016$ & $  0.246$ & $  0.420$ & $  0.527$ & $  0.281$ & $  0.401$ & $  0.543$ & $  0.187$ & $  0.353$ & $  0.004$ & $  0.004$ & $  0.160$ & $  0.174$ & $  0.191$ & $  0.140$ & $  0.097$ & $  0.091$ & $  0.112$ & $  0.003$ & $  0.002$ & $  0.468$ & $  0.395$ & $  0.325$ & $  0.234$ \\
M85-GC11 & $  0.035$ & $  0.043$ & $  0.640$ & $  0.968$ & $  1.243$ & $  0.570$ & $  0.831$ & $  1.124$ & $  0.381$ & $  0.701$ & $  0.007$ & $  0.008$ & $  0.300$ & $  0.307$ & $  0.340$ & $  0.243$ & $  0.176$ & $  0.167$ & $  0.207$ & $  0.005$ & $  0.003$ & $  1.102$ & $  0.997$ & $  1.012$ & $  0.580$ \\
M85-GC12 & $  0.021$ & $  0.027$ & $  0.499$ & $  0.737$ & $  1.016$ & $  0.525$ & $  0.788$ & $  0.995$ & $  0.318$ & $  0.655$ & $  0.006$ & $  0.007$ & $  0.285$ & $  0.314$ & $  0.332$ & $  0.248$ & $  0.169$ & $  0.164$ & $  0.192$ & $  0.004$ & $\cdots$ & $  0.852$ & $  0.639$ & $  0.685$ & $  0.422$ \\
M85-GC13 & $  0.026$ & $  0.034$ & $  0.485$ & $  0.803$ & $  0.935$ & $  0.559$ & $  0.883$ & $  1.275$ & $  0.378$ & $  0.671$ & $  0.007$ & $  0.007$ & $  0.281$ & $  0.304$ & $  0.342$ & $  0.248$ & $  0.164$ & $  0.163$ & $  0.208$ & $  0.005$ & $  0.004$ & $  0.954$ & $  0.950$ & $  0.741$ & $  0.523$ \\
M85-GC14 & $  0.031$ & $  0.039$ & $  0.493$ & $  1.378$ & $  1.024$ & $  0.553$ & $  0.786$ & $  1.108$ & $  0.364$ & $  0.715$ & $  0.007$ & $  0.008$ & $  0.313$ & $  0.314$ & $  0.352$ & $  0.260$ & $  0.178$ & $  0.167$ & $  0.209$ & $  0.005$ & $  0.004$ & $  0.913$ & $  0.839$ & $  0.717$ & $  0.545$ \\
M85-GC15 & $  0.017$ & $  0.020$ & $  0.342$ & $  0.533$ & $  0.726$ & $  0.379$ & $  0.526$ & $  0.776$ & $  0.262$ & $  0.489$ & $  0.005$ & $  0.005$ & $  0.215$ & $  0.233$ & $  0.264$ & $  0.187$ & $  0.130$ & $  0.126$ & $  0.159$ & $  0.003$ & $  0.003$ & $  0.612$ & $  0.523$ & $  0.404$ & $  0.319$ \\
M85-GC16 & $  0.030$ & $  0.037$ & $  0.469$ & $  0.869$ & $  0.974$ & $  0.536$ & $  0.936$ & $  1.096$ & $  0.376$ & $  0.669$ & $  0.007$ & $  0.008$ & $  0.285$ & $  0.312$ & $  0.331$ & $  0.254$ & $  0.174$ & $  0.162$ & $  0.198$ & $  0.005$ & $  0.004$ & $  0.895$ & $  0.882$ & $  0.663$ & $  0.489$ \\
M85-GC17 & $  0.053$ & $  0.063$ & $  0.928$ & $  1.579$ & $  1.582$ & $  0.974$ & $  1.395$ & $  1.688$ & $  0.558$ & $  0.869$ & $  0.009$ & $  0.011$ & $  0.408$ & $  0.447$ & $  0.473$ & $  0.317$ & $  0.226$ & $  0.207$ & $  0.261$ & $  0.006$ & $  0.005$ & $  1.563$ & $  1.549$ & $  1.103$ & $  0.915$ \\
M85-GC18 & $  0.055$ & $  0.074$ & $  1.227$ & $  1.365$ & $  2.752$ & $  1.144$ & $  1.167$ & $  2.019$ & $  0.545$ & $  1.045$ & $  0.010$ & $  0.012$ & $  0.427$ & $  0.458$ & $  0.483$ & $  0.375$ & $  0.250$ & $  0.240$ & $  0.302$ & $  0.007$ & $  0.005$ & $  2.417$ & $  1.541$ & $  1.505$ & $  1.003$ \\
M85-GC19 & $  0.037$ & $  0.046$ & $  0.684$ & $  1.340$ & $  2.200$ & $  0.970$ & $  1.276$ & $  1.724$ & $  0.508$ & $  0.901$ & $  0.009$ & $  0.010$ & $  0.415$ & $  0.464$ & $  0.482$ & $  0.355$ & $  0.239$ & $  0.228$ & $  0.294$ & $  0.006$ & $  0.005$ & $  1.049$ & $  1.284$ & $  0.837$ & $  0.905$ \\
M85-GC20 & $  0.050$ & $  0.064$ & $  1.080$ & $  1.725$ & $  2.290$ & $  1.363$ & $  1.746$ & $  2.249$ & $  0.724$ & $  1.121$ & $  0.012$ & $  0.013$ & $  0.507$ & $  0.534$ & $  0.568$ & $  0.413$ & $  0.288$ & $  0.286$ & $  0.313$ & $  0.007$ & $\cdots$ & $  2.496$ & $  1.573$ & $  1.794$ & $  0.947$ \\
M85-HCC1 & $  0.006$ & $  0.008$ & $  0.116$ & $  0.201$ & $  0.278$ & $  0.145$ & $  0.207$ & $  0.281$ & $  0.098$ & $  0.189$ & $  0.002$ & $  0.002$ & $  0.087$ & $  0.091$ & $  0.099$ & $  0.073$ & $  0.053$ & $  0.050$ & $  0.061$ & $  0.001$ & $  0.001$ & $  0.224$ & $  0.200$ & $  0.156$ & $  0.123$ \\
M85-Nucleus & $  0.001$ & $  0.001$ & $  0.015$ & $  0.025$ & $  0.036$ & $  0.019$ & $  0.027$ & $  0.037$ & $  0.013$ & $  0.025$ & $  <0.001$ & $  <0.001$ & $  0.012$ & $  0.012$ & $  0.013$ & $  0.010$ & $  0.007$ & $  0.007$ & $  0.008$ & $  <0.001$ & $  <0.001$ & $  0.030$ & $  0.028$ & $  0.020$ & $  0.017$ \\
\enddata
%\tablecomments{$^{a}$ The number of GC candidates assigned to fiber.}
%% Text for table notes should follow after the \enddata but before
%% the \end{deluxetable}. Make sure there is at least one \tablenotemark
%% in the table for each \tablenotetext.
%\tablenotetext{a}{CFHT/MegaCam AB magnitudes.}
\end{deluxetable*}
\end{turnpage}

	\subsection{Full Spectrum Fitting}

We also used the spectrum fitting code ULySS\footnote{http://ulyss.univ-lyon1.fr} \citep{kol09} to derive the ages and [Fe/H] of the M85 GCs.
\citet{kol08} presented the full spectrum fitting results from various combinations of SSP models and stellar libraries.
They showed that the ages and metallicities of the GCs derived from the full spectrum fitting are consistent with those from the isochrone-fitting method based on photometry of resolved stars in the GCs.
Recently, several studies have used the ULySS code to obtain stellar population parameters of GCs in nearby galaxies (e.g. Sharina et al. 2010, Cezario et al. 2013).

We fit the GMOS spectra of the 20 GCs in M85, adopting the SSP models of \citet{vaz10} computed using the MILES stellar library \citep{san06} and the Salpeter IMF.
The SSP models cover the optical spectral range of 3540.5--7409.6 \rm{\AA} at a resolution of FWHM $\sim$ 2.3 \rm{\AA}.
The ages of the SSP models range from 63 Myr to 18 Gyr and the metallicities [M/H] from --2.32 to +0.22 dex.

\clearpage

\citet{liu13} presented the effect of the signal-to-noise ratios of the spectra on the ULySS full spectrum fitting results. They found that ULySS fitting results are not reliable for the spectra with signal-to-noise ratios lower than 25. Therefore, we adopted a wavelength range with the signal-to-noise ratio of the spectra higher than 25 for the fitting. We used the wavelength range of about 5000--5800 \rm{\AA} for most target spectra.
{\bf Figure \ref{fig:fitres}} shows an example of the full spectrum fitting result for M85-GC10.
The residuals from the best fit are comparable to the 1$\sigma$ error residuals on average.
We ran 500 Monte Carlo simulations for each spectrum and determined the ages and [Fe/H] of the GCs by adopting the mean values of the 500 simulation results.
The uncertainties of the parameters correspond to the 68\% (1$\sigma$) confidence intervals of the simulation results.
The ages and metallicities of the M85 GCs, M85-HCC1, and the M85 nucleus derived using the three methods are listed in {\bf Table \ref{tab:age.zh.afe}}.

\begin{deluxetable*}{l c c c c c c c c c c}
%\tabletypesize{\scriptsize}
\setlength{\tabcolsep}{0.05in}
\tablecaption{Age, Metallicity, and [$\alpha$/Fe] of M85 GCs, M85-HCC1, and the M85 Nucleus \label{tab:age.zh.afe}}
\tablewidth{\textwidth}
\tablehead{
\colhead{ID} & \colhead{} & \colhead{Age$_{\rm grid}$} & \colhead{[Fe/H]$_{\rm grid}^{\rm [MgFe]'}$} & \colhead{[Fe/H]$_{\rm grid}^{\rm Mg_2}$} & \colhead{[$\alpha$/Fe]$_{\rm grid}$} & \colhead{Age$_{\chi^2}$} & \colhead{[Fe/H]$_{\chi^2}$} & \colhead{[$\alpha$/Fe]$_{\chi^2}$} & \colhead{Age$_{\rm ULySS}$} & \colhead{[Fe/H]$_{\rm ULySS}$} \\
\colhead{} & \colhead{} & \colhead{(Gyr)} & \colhead{(dex)} & \colhead{(dex)} & \colhead{(dex)} & \colhead{(Gyr)} & \colhead{(dex)} & \colhead{(dex)} & \colhead{(Gyr)} & \colhead{(dex)}
}
\startdata
M85-GC01 & * & $14.9 \pm <0.1$ & $0.10 \pm 0.08$ & $0.02 \pm 0.06$ & $0.00 \pm 0.05$ & $15.0^{+0.1}_{-2.3}$ & $-0.08^{+0.14}_{-0.15}$ & $0.20 \pm 0.13$ & $4.3^{+0.3}_{-0.8}$ & $0.21 \pm 0.01$ \\
M85-GC02 & & $14.0 \pm 1.1$ & $-0.97 \pm 0.14$ & $-1.30 \pm 0.12$ & $0.20 \pm 0.12$ & $15.0^{+<0.1}_{-5.6}$ & $-1.83^{+0.10}_{-0.63}$ & $0.47^{+0.03}_{-0.61}$ & $9.9^{+7.9}_{-5.8}$ & $-1.07^{+0.09}_{-0.10}$ \\
M85-GC03 & & $8.3 \pm 2.3$ & $-2.02 \pm 0.21$ & $-1.08 \pm 0.21$ & $0.49 \pm 0.02$ & $4.9^{+3.3}_{-2.0}$ & $-0.59^{+0.09}_{-0.11}$ & $0.50^{+<0.01}_{-0.07}$ & $5.9^{+11.9}_{-5.7}$ & $-1.07^{+0.17}_{-0.20}$ \\
M85-GC04 & & $3.7 \pm 1.0$ & $-0.73 \pm 0.17$ & $-0.86 \pm 0.11$ & $-0.01 \pm 0.11$ & $6.0^{+3.4}_{-2.2}$ & $-1.36^{+0.41}_{-0.44}$ & $0.08^{+0.42}_{-0.38}$ & 5.7$^{+2.1}_{-2.7}$ & $-1.28^{+0.25}_{-0.34}$ \\
M85-GC05 & * & $\cdots$ & $\cdots$ & $0.09 \pm 0.08$ & $-0.30 \pm <0.01$ & $12.0^{+2.5}_{-2.1}$ & $-0.13^{+0.15}_{-0.13}$ & $-0.21^{+0.15}_{-0.09}$ & $1.8^{+0.5}_{-0.3}$ & $0.14^{+0.08}_{-0.18}$ \\
M85-GC06 & & $14.0 \pm 0.5$ & $-0.71 \pm 0.08$ & $-0.74 \pm 0.06$ & $0.22 \pm 0.05$ & $15.0^{+<0.1}_{-2.0}$ & $-0.90^{+0.15}_{-0.18}$ & $0.37^{+0.13}_{-0.17}$ & $12.3^{+1.9}_{-1.6}$ & $-0.66^{+0.05}_{-0.04}$ \\
M85-GC07 & & $14.0 \pm 1.0$ & $-2.09 \pm 0.20$ & $-2.71 \pm 0.37$ & $0.49 \pm <0.01$ & $15.0^{+<0.1}_{-4.1}$ & $-1.31^{+0.21}_{-0.30}$ & $0.30^{+0.20}_{-0.30}$ & $14.4^{+3.4}_{-5.4}$ & $-1.15^{+0.10}_{-0.09}$ \\
M85-GC08 & & $1.4 \pm 0.1$ & $0.63 \pm 0.10$ & $0.26 \pm 0.03$ & $0.00 \pm 0.03$ & $1.3 \pm 0.1$ & $0.42^{+0.07}_{-0.13}$ & $0.27 \pm 0.07$ & $2.5 \pm 0.1$ & $0.22 \pm <0.01$ \\
M85-GC09 & & $6.2 \pm 0.6$ & $-2.63 \pm 0.11$ & $-1.56 \pm 0.13$ & $0.40 \pm 0.10$ & $3.3^{+2.3}_{-1.5}$ & $-1.46^{+0.14}_{-0.40}$ & $0.49^{+0.01}_{-0.40}$ & $3.3^{+2.2}_{-1.0}$ & $-0.33 \pm 0.10$ \\
M85-GC10 & & $1.3 \pm 0.1$ & $0.29 \pm 0.12$ & $0.12 \pm 0.03$ & $0.04 \pm 0.03$ & $1.9 \pm <0.1$ & $-0.23^{+0.07}_{-0.06}$ & 0.28$^{+0.07}_{-0.06}$ & $1.9^{+0.2}_{-0.3}$ & $0.19^{+0.03}_{-0.02}$ \\
M85-GC11 & & $14.9 \pm <0.1$ & $-0.63 \pm 0.05$ & $-0.69 \pm 0.05$ & $0.48 \pm 0.03$ & $15.0^{+<0.1}_{-2.1}$ & $-0.75^{+0.04}_{-0.09}$ & $0.50^{+<0.01}_{-0.09}$ & $16.7^{+1.1}_{-0.8}$ & $-0.45 \pm 0.03$ \\
M85-GC12 & & $\cdots$ & $\cdots$ & $-2.71 \pm <0.01$ & $0.49 \pm <0.01$ & $13.0^{+2.0}_{-1.8}$ & $-2.19^{+0.13}_{-0.24}$ & $0.50^{+<0.01}_{-0.22}$ & $8.2^{+3.0}_{-3.9}$ & $-2.09^{+0.18}_{-0.17}$ \\
M85-GC13 & & $14.9 \pm <0.1$ & $0.21 \pm 0.05$ & $-0.60 \pm 0.05$ & $-0.28 \pm 0.03$ & $15.0^{+<0.1}_{-2.6}$ & $-0.61^{+0.19}_{-0.12}$ & $-0.22^{+0.18}_{-0.08}$ & $15.1^{+2.7}_{-2.9}$ & $-1.27^{+0.09}_{-0.07}$ \\
M85-GC14 & * & $1.7 \pm 0.2$ & $0.31 \pm 0.14$ & $0.24 \pm 0.08$ & $-0.30 \pm 0.02$ & $1.4 \pm 0.1$ & $-0.07^{+0.17}_{-0.21}$ & $0.06^{+0.16}_{-0.15}$ & $16.2^{+1.5}_{-2.2}$ & $-0.94^{+0.04}_{-0.05}$ \\
M85-GC15 & & $14.0 \pm 0.5$ & $-2.30 \pm 0.17$ & $-2.71 \pm <0.01$ & $0.49 \pm <0.01$ & $13.0^{+2.0}_{-1.6}$ & $-1.65^{+0.30}_{-0.33}$ & $0.20^{+0.30}_{-0.33}$ & $10.3^{+1.6}_{-1.1}$ & $-1.40 \pm 0.09$ \\
M85-GC16 & & $2.1 \pm 0.2$ & $0.92 \pm 0.03$ & $0.45 \pm 0.04$ & $-0.29 \pm 0.03$ & $1.6^{+0.2}_{-0.1}$ & $0.28^{+0.13}_{-0.09}$ & $0.07 \pm 0.09$ & $1.6^{+<0.1}_{-0.1}$ & $0.22 \pm <0.01$ \\
M85-GC17 & * & $14.9 \pm 0.5$ & $-0.79 \pm 0.09$ & $-1.27 \pm 0.16$ & $0.49 \pm 0.07$ & $8.5^{+2.4}_{-2.5}$ & $-0.08^{+0.18}_{-0.15}$ & $-0.17^{+0.17}_{-0.13}$ & $5.7^{+1.8}_{-1.7}$ & $-0.12 \pm 0.09$ \\
M85-GC18 & & $1.1 \pm 0.1$ & $0.73 \pm 0.09$ & $0.44 \pm 0.08$ & $-0.13 \pm 0.05$ & $1.5 \pm 0.3$ & $-0.05 \pm 0.18$ & $0.21 \pm 0.17$ & $1.7^{+0.2}_{-0.4}$ & $0.17^{+0.05}_{-0.01}$ \\
M85-GC19 & & $14.9 \pm 0.7$ & $-2.55 \pm 0.25$ & $-2.19 \pm 0.23$ & $0.36 \pm 0.19$ & $15.0^{+<0.1}_{-7.4}$ & $-2.40^{+0.27}_{-0.80}$ & $0.50^{+<0.01}_{-0.80}$ & $6.0^{+2.8}_{-2.7}$ & $-1.22^{+0.13}_{-0.12}$ \\
M85-GC20 & & $14.8 \pm 1.1$ & $-0.18 \pm 0.11$ & $-0.74 \pm 0.10$ & $-0.24 \pm 0.09$ & $13.9^{+1.1}_{-5.0}$ & $-0.73^{+0.39}_{-0.29}$ & $-0.06^{+0.37}_{-0.24}$ & $16.9^{+0.8}_{-0.3}$ & $-0.18 \pm 0.07$ \\
M85-HCC1 & & $1.4 \pm 0.1$ & $0.20 \pm 0.03$ & $0.18 \pm 0.01$ & $0.01 \pm 0.01$ & $1.6 \pm <0.1$ & $0.10 \pm 0.03$ & $0.17 \pm 0.03$ & $2.0 \pm <0.1$ &  $0.22^{+<0.01}_{-0.01}$ \\
M85-Nucleus & & $1.8 \pm <0.1$ & $0.14 \pm 0.01$ & $0.25 \pm 0.01$ & $0.25 \pm 0.01$ & $2.6 \pm <0.1$ & $0.11^{+0.01}_{-<0.01}$ & $0.23^{+0.01}_{-<0.01}$ & $3.8 \pm 0.1$ & $0.22 \pm <0.01$ \\
\enddata
%\tablecomments{$^{a}$ The number of GC candidates assigned to fiber.}
%% Text for table notes should follow after the \enddata but before
%% the \end{deluxetable}. Make sure there is at least one \tablenotemark
%% in the table for each \tablenotetext.
\tablenotetext{}{Asterisks next to the ID indicate the outliers in {\bf Figure \ref{fig:comp_age_afe}(a) and (c)}.}
\end{deluxetable*}

	\subsection{Comparison of Parameters}

	\begin{figure}
\epsscale{1}
\includegraphics[width=\columnwidth]{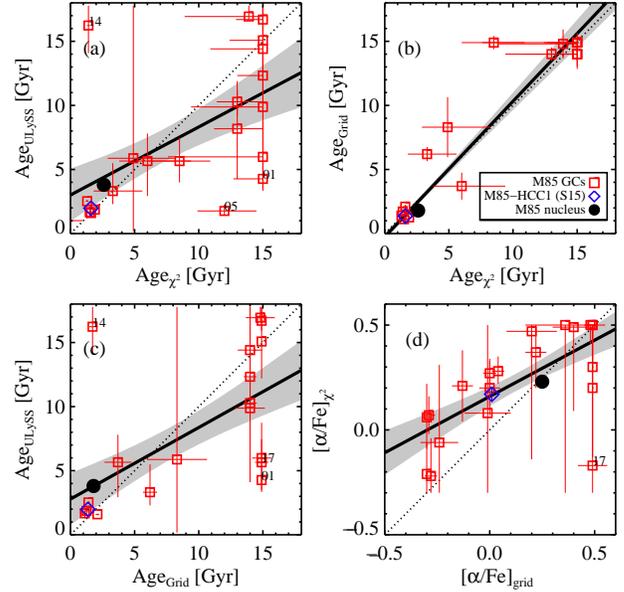}
\caption{(a) Comparison between the ages from the ULySS fitting and $\chi^2$ minimization method. 
(b) Comparison between the ages from the Lick index grid method and $\chi^2$ minimization method.
(c) Comparison between the ages from the ULySS fitting and Lick index grid method.
(d) Comparison between the [$\alpha$/Fe] from the $\chi^2$ minimization method and Lick index grid method.
The red squares, blue diamond, and filled circle represent the M85 GCs, M85-HCC1, and the M85 nucleus, respectively.
The dotted line denotes the one-to-one relation. 
The solid lines and shaded regions represent the least-squares fit with the data and 1$\sigma$ uncertainties, respectively.
The IDs of the outliers from the one-to-one relation are marked in the panels.
\label{fig:comp_age_afe}}
	\end{figure}	

	\begin{figure*}
\epsscale{1}
\includegraphics[width=\textwidth]{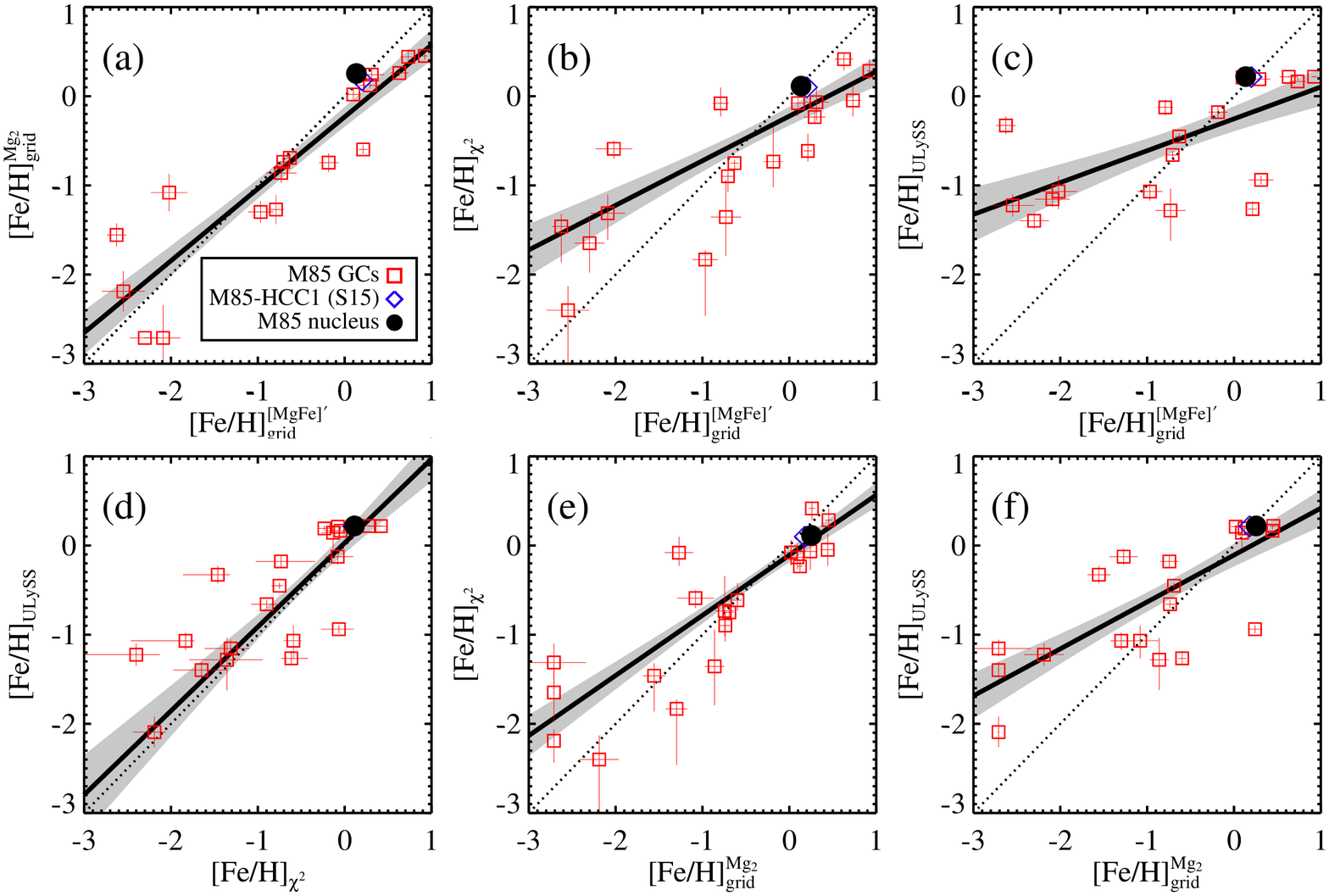}
\caption{Comparison of [Fe/H] for the M85 GCs measured using different methods: (a) Mg$_2$ grid vs. [MgFe]$'$ grid, (b) $\chi^2$ minimization method vs. [MgFe]$'$ grid, (c) full spectrum fitting vs. [MgFe]$'$ grid, (d) full spectrum fitting vs. $\chi^2$ minimization method, (e) $\chi^2$ minimization method vs. Mg$_2$ grid, and (f) full spectrum fitting vs. Mg$_2$ grid.
The symbols and lines are the same as {\bf Figure \ref{fig:comp_age_afe}}.
\label{fig:comp_feh}}
	\end{figure*}

We compared the ages, metallicities, and [$\alpha$/Fe] derived using three different methods.
There are some caveats for each method described in Sections 4.1 and 4.2.
For the Lick index grid method, the ages of the GCs are only derived from the H$\beta$--[MgFe]$'$ grid.
We could not measure the ages of GC05 and GC12 because their [MgFe]$'$ indices are not real numbers.
The dependence on H$\beta$ in determining the ages of GCs may be problematic because the H$\beta$ line strength is strongly affected by the horizontal branch morphology \citep{bur84,db95,cbc03}.
The blue horizontal branch stars in GCs strengthen the H$\beta$ line, and consequently, GCs that have a stronger blue horizontal branch morphology result in younger ages.
In addition, the $\alpha$-element abundance is hard to distinguished at low metallicities as shown in {\bf Figure \ref{fig:lick_diagram}(b)}.
As mentioned in Section 4.1.1, if the GCs do not fall on the grid, we cannot determine their ages and metallicities directly. In that case, we adopt the grid boundary values.

The $\chi^2$ minimization method uses more indices than the Lick index grid method, but there may be cases in which the fitting results depend heavily on only a few indices.
We checked the stability of the fitting results by further excluding the Lick indices used for the fitting one by one.
The fitting results for GC01 and GC05 are unstable if Mg and Fe lines are rejected.
For example, the age and [Fe/H] of GC01 derived from the $\chi^2$ minimization method are 15 Gyr and --0.06 dex.
However, if the Fe5015 index is excluded in the fitting process, they are derived to be $\sim$ 2 Gyr and +0.70 dex.
This large difference caused by the exclusion of one line indicates that the fitting is unstable, not converging.
In addition, the ages of five GCs (GC02, GC03, GC17, GC18, and GC20) strongly depend on the H$\beta$ index.
For these five GCs, other Balmer lines, H$\gamma$ and H$\delta$, are excluded in the fitting process because of their low signal-to-noise ratios.
The absence of Balmer lines with a large number of measured Lick indices for the fitting process of the $\chi^2$ minimization method affects minimally the derived ages, even though the Balmer lines are strong and age-sensitive lines \citep{pro04}.
However, for these GCs, the number of indices used in the fitting process is too small to exclude the H$\beta$ line.
We deal with the derived parameters of these seven GCs carefully in further comparison.

These two methods based on Lick indices may give inaccurate properties because our values of the Lick indices are not calibrated to the Lick standard system.
The full spectrum fitting is not affected by this calibration issue, but we cannot derive the $\alpha$-element abundances using the Vazdekis models in this method.

{\bf Figure \ref{fig:comp_age_afe}} shows a comparison of the ages and [$\alpha$/Fe] obtained using different methods. The ages based on the Lick index grid method and the $\chi^2$ minimization method are consistent within uncertainties as shown in {\bf Figure \ref{fig:comp_age_afe}(b)}. We derive a linear least-squares fit: Age(grid) = 1.0($\pm$0.1)Age($\chi^2$) -- 0.2($\pm$0.3) with RMSE = 1.9 Gyr. {\bf Figure \ref{fig:comp_age_afe}(a) and (c)} show a comparison of the ages obtained from the Lick indices and the full spectrum fitting. The linear least-squares fitting results are as follows: Age(ULySS) = 0.5($\pm$0.2)Age($\chi^2$) + 3.0($\pm$2.0) with RMSE = 4.4 Gyr and Age(ULySS) = 0.6($\pm$0.2)Age(grid) + 2.9($\pm$2.1) with RMSE = 4.4 Gyr. It is noted that there are four outliers showing about 3$\sigma$ deviations (GC01, GC05, GC14, and GC17). Among them, the values for GC01, GC05, and GC17 are based on unstable $\chi^2$ fitting results as previously mentioned. We cannot find any possible clues to explain the other outlier, GC14. {\bf Figure \ref{fig:comp_age_afe}(d)} shows a comparison of the $\alpha$-element abundances derived from the Lick index grid method and the $\chi^2$ minimization method. We find a loose correlation between the two within uncertainties: [$\alpha$/Fe]($\chi^2$) = 0.54($\pm$0.15)[$\alpha$/Fe](grid) + 0.16($\pm$0.04) with RMSE = 0.18 dex. The [$\alpha$/Fe] obtained from the $\chi^2$ minimization method are larger than those obtained from the Lick index grid method on average.

{\bf Figure \ref{fig:comp_feh}} shows a comparison of the metallicities obtained using different methods. We calculate the iron abundance of the GCs based on the Lick indices using the relation presented by \citet{tmb03}: [Fe/H] = [Z/H] -- 0.94[$\alpha$/Fe]. {\bf Figure \ref{fig:comp_feh}(a)} shows that the [Fe/H] values derived from the Lick index grid method agree well. We derive a linear least-squares fit: [Fe/H](grid,Mg$_2$) = 0.80($\pm$0.09)[Fe/H](grid,[MgFe]$'$) -- 0.23($\pm$0.11) with RMSE = 0.40 dex. {\bf Figure \ref{fig:comp_feh}(b) and (e)} show a comparison of [Fe/H] derived from the $\chi^2$ minimization method and the grid method. We derive linear least-squares fits: [Fe/H]($\chi^2$) = 0.50($\pm$0.10)[Fe/H](grid,[MgFe]$'$) -- 0.22($\pm$0.11) with RMSE = 0.47 dex and [Fe/H]($\chi^2$) = 0.67($\pm$0.09)[Fe/H](grid,Mg$_2$) -- 0.11($\pm$0.08) with RMSE = 0.41 dex. {\bf Figure \ref{fig:comp_feh}(c) and (f)} show a comparison between the [Fe/H] values derived from the full spectrum fitting and the grid method. The linear least-squares fits for these data are as follows: [Fe/H](ULySS) = 0.36($\pm$0.11)[Fe/H](grid,[MgFe]$'$) -- 0.25($\pm$0.14) with RMSE = 0.47 dex and [Fe/H](ULySS) = 0.53($\pm$0.10)[Fe/H](grid,Mg$_2$) -- 0.11($\pm$0.13) with RMSE = 0.43 dex. The [Fe/H] values derived between the full spectrum fitting and the grid method show a big difference for metal-poor GCs ([Fe/H] $<$ --2). This difference is weakened between the [Fe/H] values from the full spectrum fitting and the $\chi^2$ minimization method.

As a result, we find that the ages and metallicities derived from the Lick indices and the full spectrum fitting follow the same trend, but there are inconsistencies for several GCs. These differences might be due to the calibration problem on the Lick indices in addition to the different stellar population models we used for the different methods. Therefore, only the results obtained from the full spectrum fitting will be considered in the following analysis.

We compare the ages and metallicities of the M85 nucleus and M85-HCC1 with those derived in previous spectroscopic studies.
\citet{ffi96} suggested that the age of the M85 nucleus is about 3 Gyr, and \citet{tf02} derived the age and [Fe/H] of M85 as 1.6 Gyr and 0.44 dex.
We measure the age and [Fe/H] of the M85 nucleus as 3.8 $\pm$ 0.1 Gyr and 0.22 $\pm <$0.01 dex.
All of the results agree that the M85 nucleus has an intermediate-age stellar population with a supersolar metallicity.

\citet{san15} presented the values of the age, [Fe/H], and [Mg/Fe] ratio derived from the SDSS spectrum of M85-HCC1.
The values of the parameters are 3.0 $\pm$ 0.4 Gyr, --0.06 $\pm$ 0.07 dex, and 0.05$\pm$ 0.13 dex, respectively.
We derive the age and [Fe/H] of M85-HCC1 to be 2.0 $\pm <$0.1 Gyr and 0.22$^{+<0.01}_{-0.01}$ dex.
%Thus our values are consistent with those in \citet{san15}.
It is interesting that M85-HCC1 and the M85 nucleus have similar ages and metallicities, which indicates that M85-HCC1 might have formed with the recent star formation in the M85 nucleus.
%Despite slight differences, we conclude that the M85 nucleus  and M85-HCC1 have young ages and solar metallicities, consistent with the results of previous studies.

	\subsection{Age and Metallicity Distribution of M85 GCs}

	\begin{figure*}[hbt]
\epsscale{1}
\includegraphics[width=\textwidth]{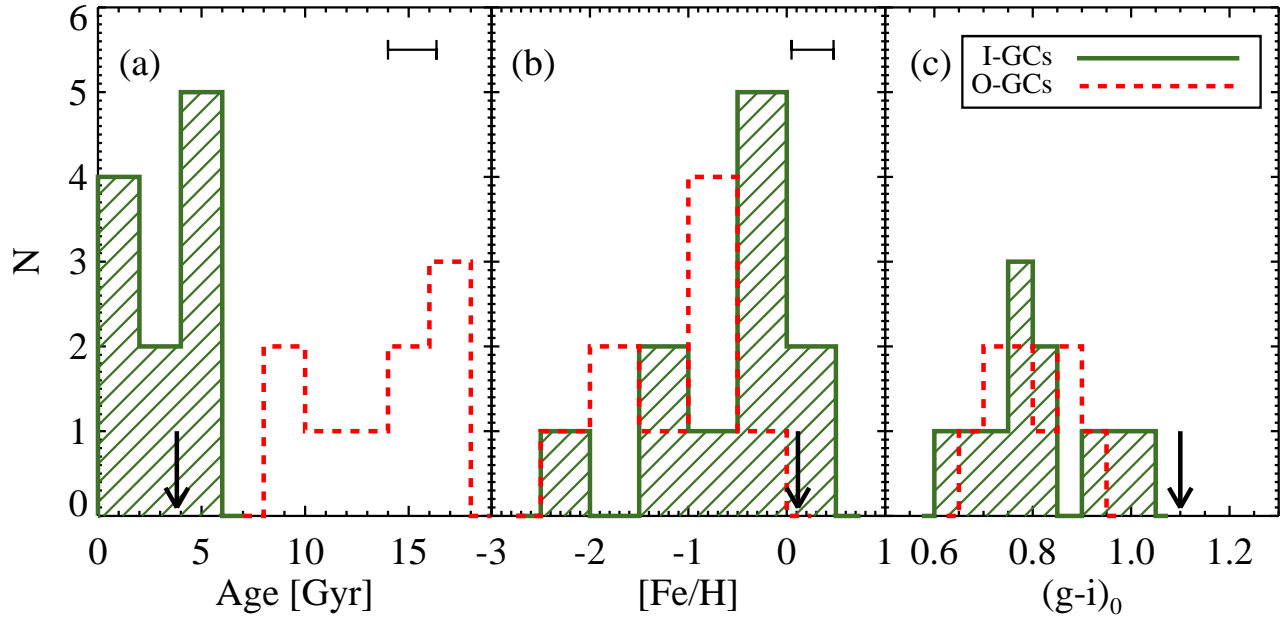}
\caption{(a) Age, (b) [Fe/H], and (c) $(g-i)_0$ color distributions of the GCs.
The hatched solid and dashed histograms represent the intermediate-age and old GCs, respectively. 
The arrows indicate the age, [Fe/H], and $(g-i)_0$ color of the M85 nucleus derived in this study.
The error bars in the upper-right corner in panels (a) and (b) indicate the mean errors of the ages and [Fe/H], respectively. 
\label{fig:measure_dist1}}	
	\end{figure*}	
		
{\bf Figure \ref{fig:measure_dist1}} shows the age, [Fe/H], and $(g-i)_0$ color distributions of the M85 GCs confirmed in this study.
We find two groups of M85 GCs based on their ages: intermediate-age GCs with ages of $<$ 8 Gyr and old GCs with ages of $>$ 8 Gyr.
The number of intermediate-age and old GCs are 11 and 9, respectively.
The mean age of the intermediate-age GCs is about 3.7 Gyr with a standard deviation of 1.9 Gyr, which is similar to the stellar age of the M85 nucleus ($\sim$ 3.8 Gyr) within 1$\sigma$ uncertainty, while the old GCs have a mean age of 13.3 Gyr with a standard deviation of 3.3 Gyr.
The intermediate-age GCs have, on average,  higher H$\beta$ indices than the old GCs (see {\bf Figure \ref{fig:lick_diagram}(a)}). However, there are some outliers. Three of the intermediate-age GCs (M85-GC01, M85-GC17, and M85-GC19) are located below the 15 Gyr line of the grid. The ages of these three GCs are estimated to be as old as 15 Gyr from the Lick index grid method. On the other hand, there is one old GC, M85-GC14, which has a H$\beta$ index as high as that of the intermediate-age GCs. The estimated age of M85-GC14 based on the Lick index grid method is 1.7 Gyr.

The fraction of intermediate-age GCs among all observed GCs is 55 $\pm$ 17\%, which is much smaller than the value, $\sim$ 85\%, estimated by \citet{tra14}.
\citet{tra14} detected a 1.8 Gyr old GC population based on optical and near-infrared photometry.
The difference may be caused by the difference of the observation coverage between the two studies.
We cover an area seven times larger than that in \citet{tra14}, but analyze a sample three times smaller than the sample in \citet{tra14}.

We investigate the [Fe/H] and $(g-i)_0$ color distributions of two GC subpopulations separately.
The intermediate-age GCs have a solar metallicity with a large dispersion, [Fe/H] = --0.26 $\pm$ 0.62.
The mean metallicity of the old GCs is [Fe/H] = --1.02 $\pm$ 0.56, which is lower than that of the intermediate-age GCs.
The intermediate-age and old GCs have a color range of $0.60 < (g-i)_0 < 1.05$ and $0.65 < (g-i)_0 < 0.95$, respectively, with similar mean colors of $(g-i)_0 \sim 0.8$ (see {\bf Figure 9(c)}). The two GC subpopulations do not show a significant difference in their color distributions.
Old GCs in massive early-type galaxies generally show a color bimodality, and the GC populations are divided into blue and red subpopulations with a division color criterion of $(g-i)_0$ = 0.8 \citep{dur14}. We expect that the intermediate-age GCs could weaken the color bimodality of the old GCs.
However, there is no clear color bimodality for the old GCs in M85, and the mean $(g-i)_0$ color of the intermediate-age GCs is redder than that of the old GCs.
This could be due to small number statistics, so it needs to be confirmed with larger samples in the future. 
In addition, we do not find any clear radial gradients of age or metallicity for the GCs in M85.

	\section{K\MakeLowercase{inematics of the} GC s\MakeLowercase{ystem in} M85}

	\begin{figure}
\epsscale{0.9}
\includegraphics[width=\columnwidth]{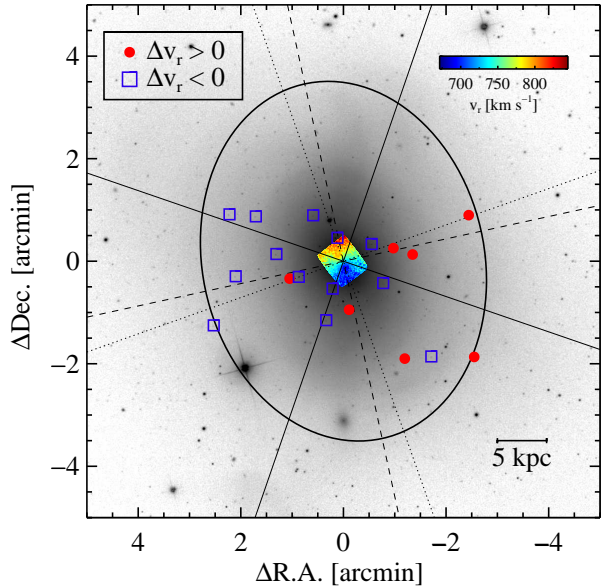}
\caption{Spatial distribution of the 20 M85 GCs confirmed in this study.
The filled circles and open squares represent the GCs with radial velocity higher and lower than that of the M85 nucleus, respectively.
The solid line ellipse shows the stellar extent of M85 with a major axis of $7'$. 
The solid lines represent the kinematic major/minor axes of the GC system derived in \S5.
The stellar velocity field derived from the SAURON project \citep{ems07} is overlaid on the SDSS image, the which velocity scales of which are displayed on the right corner.
The photometric and kinematic position angles of the stellar light of M85 is taken from Krajnovi{\'c} et al. (2011; PA$_{\rm phot}$ = 12$\fdg$3 $\pm$ 11$\fdg$0 and PA$_{\rm kin}$ = 19$\fdg$5 $\pm$ 4$\fdg$8).
The dashed lines and dotted lines represent the photometric major/minor axes and kinematic major/minor axes of the stars, respectively.
\label{fig:v_spatial}}
	\end{figure}

	\begin{figure}
\epsscale{0.9}
\includegraphics[width=\columnwidth]{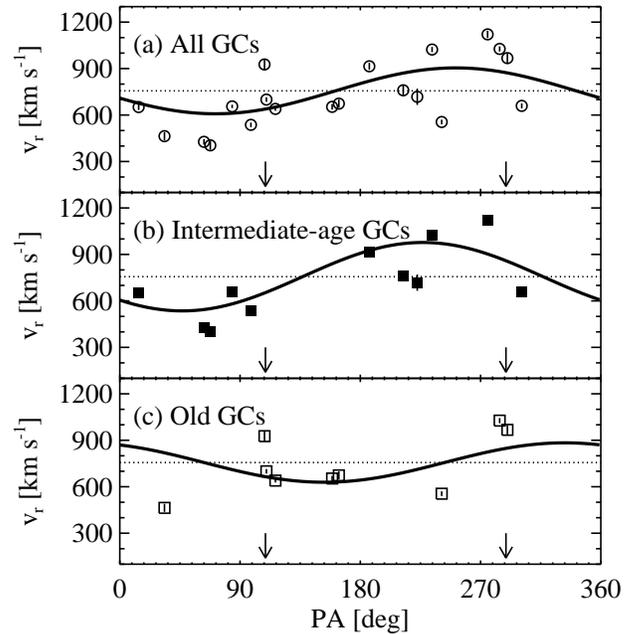}
\caption{Radial velocities of (a) all GCs, (b) intermediate-age GCs ($<$ 8 Gyr), and (c) old GCs ($>$ 8 Gyr) as a function of position angle.
The solid line curves and dotted lines represent the best-fit rotation curve and the radial velocity of the M85 nucleus derived in this study, respectively.
The vertical arrows show the photometric minor axis of M85 \citep{kra11}.
\label{fig:pa_vr}}	
	\end{figure}

	\begin{figure}
\epsscale{1}
\includegraphics[width=\columnwidth]{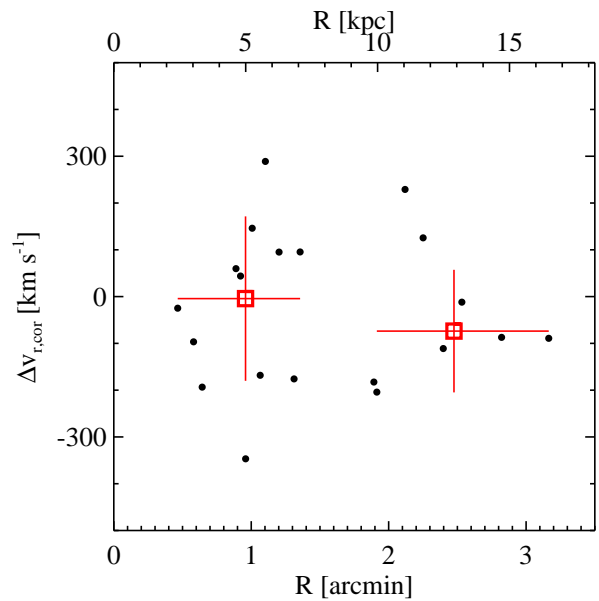}
\caption{Rotation-corrected radial velocity of the 20 GCs relative to the radial velocity of the M85 nucleus vs. galactocentric distance.
The open squares represent the mean radial velocities of the GCs in each radial bin.
The vertical error bars indicate the radial velocity dispersions of the GCs in each bin.
\label{fig:rad_dvr}}	
	\end{figure}

The mean radial velocity of the M85 GCs measured in this study is $\overline{v_r} = 724^{+45}_{-46}$ \kmsend, which is similar to that of the M85 nucleus (\vrad = 756 $\pm$ 8 \kmsend).
Considering this and the Gaussian distribution of their radial velocity, we conclude that these GCs are indeed gravitationally bound to M85.
We estimate the radial velocity dispersion and its error for the M85 GCs, $\sigma_{\rm r} = 202^{+19}_{-33}$ \kmsend.
It is slightly higher than the velocity dispersion of the M85 nucleus ($\sigma = 172$ \kmsend).

{\bf Figure \ref{fig:v_spatial}} shows the spatial distribution of the GCs confirmed in this study.
The GCs with radial velocity higher than that of the M85 nucleus are mostly located in the western region, while the others are in the eastern region.
This spatial segregation of the GCs shows that the GC system of M85 is rotating.
It is notable that the rotation signature is clear, although M85 is nearly a face-on galaxy.

The GCs rotating around a given axis in the plane of the sky have radial velocities depending sinusoidally on the position angle.
We measure the rotation amplitude and position angle of the rotation axis for the M85 GC system by fitting the data to the following function, 
\begin{displaymath}
    v_{\rm r}(\Theta)~=~v_{\rm sys} + (\Omega R)~{\rm sin}(\Theta-\Theta_0),
\end{displaymath}
where $v_{\rm sys}$ is the systemic velocity of the GC system, $\Omega R$ is the rotation amplitude, and $\Theta_0$ is the orientation of the rotation axis.
The systemic velocity of the M85 GC system is assumed to be the radial velocity of the M85 nucleus, \vrad = 756 \kmsend.
{\bf Figure \ref{fig:pa_vr}(a)} shows the radial velocities of the 20 GCs as a function of position angle with their best-fit rotation curves.
The rotation amplitude and orientation of the rotation axis of the entire GC system are derived to be $\Omega R = 148^{+67}_{-42}$ \kms and $\Theta_0 = 161^{+25}_{-18} \arcdeg$.
In addition, we correct the projected rotation amplitude by considering the inclination of M85.
The $sin~i$ factor for the inclination angle $i=39^\circ$ is about 0.63, assuming that M85 has equal major and minor axes.
Therefore, the rotation amplitude corrected for inclination, which is about 1.6 times larger than the projected one, is $\Omega R_{\rm icor} = 235^{+107}_{-66}$ \kmsend.
The position angle of the photometric major axis is about 12$\arcdeg$ \citep{kra11}, which shows a significant deviation from that of the orientation of the rotation axis.
We will discuss the difference between these position angles in Section 6.1.

In addition to the entire GC system, we measure the rotation parameters for the two GC subgroups considered above, the intermediate-age ($<$ 8 Gyr) and old ($>$ 8 Gyr) GC systems.
{\bf Figure \ref{fig:pa_vr}(b) and (c)} show the rotation curves for 11 intermediate-age GCs and 9 old GCs, respectively.
The rotation amplitudes of the intermediate-age and old GCs are $\Omega R = 221^{+62}_{-34}$ \kms and 127$^{+128}_{-13}$ \kmsend, respectively, corresponding to $\Omega R_{\rm icor} = 351^{+99}_{-54}$ \kms and 202$^{+204}_{-20}$ \kms after inclination correction.
The orientations of the rotation axes of the intermediate-age and old GCs are $\Theta_0 = 137^{+24}_{-15} \arcdeg$ and 243$^{+30}_{-78} \arcdeg$, respectively, which are also consistent with those of the entire GC system within uncertainties.

We derive the rotation-corrected radial velocities of the M85 GCs by applying the rotation curve of the entire GC system.
The rotation-corrected radial velocity dispersion of the M85 GC system is $\sigma_{\rm r,cor} = 160^{+17}_{-29}$ \kmsend, which is much smaller than the rotation amplitude of the M85 GC system.
The rotation parameter of the M85 GC system is $\Omega R_{\rm icor}/\sigma_{r,cor}$ = 1.47$^{+1.05}_{-0.48}$. We derive the values of $\Omega_{\rm icor} / \sigma_{\rm r,cor}$ = 2.41$^{+1.29}_{-0.27}$ and 1.14$^{+2.35}_{-0.09}$ for the intermediate-age GC and old GC systems, respectively. Thus, the intermediate-age GC system shows a stronger rotation than the old GC system.
We caution, however, that the kinematics of these GC subpopulations can be biased toward showing rotational support due to small number statistics. \citet{tol16} investigated the kinematics of the GC system in six Virgo early-type dwarf galaxies. They performed simulations to explore the effects of sample size on the kinematic analysis and found that the rotation amplitudes could be overestimated if the number of sample GCs is smaller than 20. Therefore, the strong rotation features of the intermediate-age and old GC systems in M85 found in this study need be checked with a larger sample size.

{\bf Figure \ref{fig:rad_dvr}} shows the rotation-corrected radial velocities of the M85 GCs relative to the radial velocity of the M85 nucleus as a function of galactocentric distance from M85.
The rotation-corrected radial velocity dispersions of the GCs in the inner and outer regions are $\sigma_{\rm r,cor} =  176^{+18}_{-45}$ \kms and $131^{+12}_{-50}$ \kmsend, respectively.
The radial velocity dispersion of the GC system appears to decrease slightly as the galactocentric distance increases, but more samples are needed to reduce its uncertainties.
The kinematics of the GC systems in M85 are summarized in {\bf Table \ref{tab:kinematics}}.
We estimate the uncertainties of the kinematic parameters corresponding to 68\% (1$\sigma$) confidence intervals.
We construct 1000 artificial data sets from the actual data for a numerical bootstrap procedure, estimate the kinematic parameters, and identify the 16th and 84th percentiles from the sorted results.
The uncertainties are defined as the differences between these values and the parameters computed by the actual data.

\begin{deluxetable*}{l c c c c c c c c}
\tabletypesize{\small}
\tablecaption{Kinematics of the M85 GC System \label{tab:kinematics}}
\tablewidth{\textwidth}
\tablehead{
\colhead{Sample} & \colhead{$N$} & \colhead{$\overline{v_r}$} & \colhead{$\sigma_r$}  & \colhead{$\Omega R$}  & \colhead{$\Omega R_{\rm icor}$} & \colhead{$\Theta_0$} & \colhead{$\sigma_{\rm r, cor}$} & \colhead{$\Omega R_{\rm icor}$/$\sigma_{\rm r, cor}$} \\
\colhead{} & \colhead{} & \colhead{(\kmsend)} & \colhead{(\kmsend)} & \colhead{(\kmsend)} & \colhead{(\kmsend)} & \colhead{(deg)} & \colhead{(\kmsend)} & \colhead{}
}
\startdata
All GCs & 20 & 724$^{+45}_{-46}$ & 202$^{+19}_{-33}$ & 148$^{+67}_{-42}$ & 235$^{+107}_{-66}$ & 161$^{+25}_{-18}$ & 160$^{+17}_{-29}$ & 1.47$^{+1.05}_{-0.48}$ \\
Intermediate-age GCs & 11 & 716$^{+68}_{-63}$ & 217$^{+24}_{-53}$ & 221$^{+62}_{-34}$ & 351$^{+99}_{-54}$ & 137$^{+24}_{-15}$ & 145$^{+11}_{-36}$ & 2.41$^{+1.29}_{-0.27}$ \\
Old GCs & 9 & 734$^{+59}_{-64}$ & 182$^{+16}_{-33}$ & 127$^{+128}_{-13}$ & 202$^{+204}_{-20}$ & 243$^{+30}_{-78}$ & 177$^{+16}_{-73}$ & 1.14$^{+2.35}_{-0.09}$ \\
\enddata
%\tablecomments{$^{a}$ The number of GC candidates assigned to fiber.}
%% Text for table notes should follow after the \enddata but before
%% the \end{deluxetable}. Make sure there is at least one \tablenotemark
%% in the table for each \tablenotetext.
%\tablenotetext{a}{CFHT/MegaCam AB magnitudes.}
\end{deluxetable*}

\section{D\MakeLowercase{iscussion}}

	\subsection{Kinematical Decoupling Between GCs and Stellar Light}

The comparison of the kinematics of the GC system with those of the stellar light in their host galaxies provides useful clues to understand the formation and evolution history of their host galaxies.
\citet{li15} investigated the kinematics of the central stellar light and the GC system in the outer region of four early-type Virgo Cluster galaxies.
They found that two of them, VCC 2000 (NGC 4660) and VCC 685 (NGC 4350), show a clear misalignment of the kinematic position angles of the stars and GC system.
These differences imply that the formation process of stars in the galaxy center is different from that of the GC system in the outer regions.
The outer region of galaxies preserves merging or accretion signatures related to its formation history. 

We compare the kinematics of the M85 GC system in this study with those of the stellar light in M85 investigated by ATLAS$^{\rm 3D}$ \citep{cap11,ems11,kra11}.
\citet{kra11} presented the kinematic maps of 260 early-type galaxies with a field of view of 33 $\times$ 41 arcsec$^2$ using the SAURON integral-field spectrograph and measured the kinematic misalignment angles.
According to their analysis, the central stellar light of M85 is dominated by ordered rotation and shows a smooth variation of the kinematic orientation within 10$\arcdeg$.
The kinematic position angle of the M85 stellar light derived by \citet{kra11} is PA$_{\rm kin}$ = 19$\fdg$5 $\pm$ 4$\fdg$8.
The kinematic position angle is defined as the angle from the north to the maximum receding part of the velocity map.
Thus, the stars in the central region of M85 rotate approximately about the photometric minor axis.
\citet{kra11} estimated the photometric position angle of M85 at a much larger scale, $R=5\arcmin-6\arcmin$.
The photometric position angle of M85 is PA$_{\rm phot}$ = 12$\fdg$3 $\pm$ 11$\fdg$0. This value is not much different from that of the kinematic position angle.
Therefore, the kinematic misalignment of stellar light in the central region of M85 is not significant.

The kinematic position angle of the M85 GCs is derived to be PA$_{\rm GCkin}$ = 251$\arcdeg$ (see \S5 and {\bf Figure \ref{fig:pa_vr}}). 
This value is 232$\arcdeg$ larger than the kinematic position angle of stellar light in M85. 
Thus, the GC system shows a rotation axis clearly distinct from that of the stars (see {\bf Figure \ref{fig:v_spatial}}).
The decoupled rotation features of the stars and GC systems in M85 implies that M85 has undergone merging events.

		\subsection{The Merging History of M85}

	\begin{figure}[hbt]
\epsscale{1}
\includegraphics[trim=60 10 20 0,clip,width=\columnwidth]{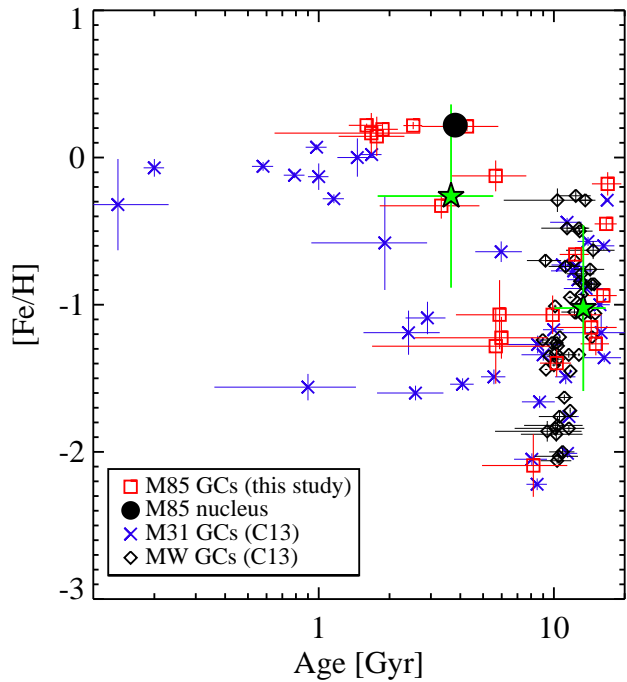}
\caption{Age--metallicity relation of the M85 GCs (red squares), Galactic GCs (black open diamonds), and M31 GCs (blue crosses) from the ULySS full spectrum fitting.
The ages and metallicities of the GCs in the Milky Way and M31 are adopted from \citet{cez13}.
The filled circle represents the age and metallicity of the M85 nucleus.
The filled stars indicate the mean ages and metallicities of the intermediate-age and old GCs, and their error bars represent the standard deviations of the ages and metallicities of two subpopulations. \label{fig:measure_dist2}}	
	\end{figure}	
	
\begin{deluxetable*}{l c c c c c c c c c c}
%\rotate
\tabletypesize{\scriptsize}
\tablecaption{Kinematics of the GC systems in early-type galaxies \label{tab:kinematics_etg}}
\tablewidth{0pt}
\tablehead{
\colhead{Galaxy} & \colhead{Type$^{a}$} & \colhead{$D^{b}$} & \colhead{$M_K^{c}$} & \colhead{$v_{\rm sys}^{d}$} & \colhead{$N_{\rm GC}$} & \colhead{$\overline{v_r}$} & \colhead{$\Omega R$}  & \colhead{$\sigma_{r, cor}$} & \colhead{$\Omega R$/$\sigma_{r, cor}$} & \colhead{Ref.$^{e}$} \\
\colhead{} & \colhead{} & \colhead{(Mpc)} & \colhead{(mag)} & \colhead{(\kmsend)} & \colhead{} & \colhead{(\kmsend)} & \colhead{(\kmsend)} & \colhead{(\kmsend)} & \colhead{}
}
\startdata
M85 & S0pec & 17.9 & --25.1 & 756 & 20 & 724$^{+45}_{-46}$ & 235$^{+107}_{-66}$ & 160$^{+17}_{-29}$ & 1.47$^{+1.05}_{-0.48}$ & This study \\
M60 & E2 & 17.3 & --25.5 & 1056 & 121 & 1073$^{+22}_{-22}$ & 141$^{+50}_{-38}$ & 217$^{+14}_{-16}$ & 0.65$^{+0.27}_{-0.22}$ & 1 \\
& & & & & 366 & $\cdots$ & 95$^{+18}_{-17}$ & 229$^{+7}_{-8}$ & 0.41 & 9 \\
M87 & cDpec & 17.2 & --25.4 & 1307 & 276 & 1333$^{+25}_{-23}$ & 172$^{+39}_{-28}$ & 399$^{+15}_{-18}$ & 0.43$^{+0.12}_{-0.09}$ & 1(2) \\
& & & & & 737 & 1336 & $\cdots$ & $\cdots$ & 0.07$^{+0.05}_{-0.04}$ & 10 \\
M49 & E2 & 17.1 & --25.8 & 997 & 263 & 973$^{+20}_{-18}$ & 54$^{+50}_{-23}$ & 321$^{+14}_{-17}$ & 0.17$^{+0.17}_{-0.08}$ & 1(3) \\
NGC 1399 & E1pec & 20.0 & --25.2 & 1442 & 435 & 1442$^{+15}_{-14}$ & 31$^{+43}_{-48}$ & 326$^{+11}_{-13}$ & 0.10$^{+0.14}_{-0.15}$ & 1(4) \\
NGC 5128 & S0pec & 4.2 & --24.2 & 541 & 210 & 536$^{+9}_{-8}$ & 30$^{+16}_{-14}$ & 129$^{+5}_{-7}$ & 0.23$^{+0.14}_{-0.13}$ & 1(5) \\	
NGC 4636 & E0-1 & 14.7 & --24.4 & 928 & 172 & 899$^{+16}_{-14}$ & 29$^{+35}_{-14}$ & 207$^{+10}_{-11}$ & 0.14$^{+0.18}_{-0.08}$ & 1(6) \\
& & & & & 238 & 949$^{+13}_{-16}$ & 37$^{+32}_{-30}$ & 226$^{+12}_{-9}$ & 0.16$^{+0.14}_{-0.14}$ & 7 \\
NGC 1407 & E0 & 26.8 & --25.4 & 1784 & 172 & 1753$^{+19}_{-18}$ & 86$^{+27}_{-35}$ & 247$^{+12}_{-13}$ & 0.35$^{+0.11}_{-0.12}$ & 7(8) \\
& & & & & 369 & 1774 & $\cdots$ & $\cdots$ & 0.17$^{+0.07}_{-0.08}$ & 10 \\
NGC 821 & E6 & 23.4 & --24.0 & 1718 & 61 & 1750 & $\cdots$ & $\cdots$ & 0.11$^{+0.18}_{-0.18}$ & 10 \\
NGC 1400 & SA0 & 26.8 & --23.8 & 558 & 72 & 612 & $\cdots$ & $\cdots$ & 0.07$^{+0.17}_{-0.13}$ & 10 \\
NGC 2768 & E6 & 21.8 & --25.4 & 1353 & 109 & 1338 & $\cdots$ & $\cdots$ & 0.39$^{+0.13}_{-0.11}$ & 10 \\
NGC 3377 & E6 & 10.9 & --22.7 & 690 & 126 & 685 & $\cdots$ & $\cdots$ & 0.19$^{+0.13}_{-0.14}$ & 10 \\
NGC 4278 & E2 & 15.6 & --23.8 & 620 & 256 & 637 & $\cdots$ & $\cdots$ & 0.17$^{+0.08}_{-0.08}$ & 10 \\
NGC 4365 & E3 & 23.3 & --25.2 & 1243 & 269 & 1210 & $\cdots$ & $\cdots$ & 0.11$^{+0.09}_{-0.10}$ & 10 \\
NGC 5846 & E0 & 24.2 & --25.0 & 1712 & 195 & 1706 & $\cdots$ & $\cdots$ & 0.02$^{+0.15}_{-0.12}$ & 10 \\
NGC 7457 & S0 & 12.9 & --22.4 & 844 & 27 & 847 & $\cdots$ & $\cdots$ & 1.68$^{+0.37}_{-0.40}$ & 10(10+11) \\
NGC 3115 & S0 & 9.4 & --24.0 & 663 & 190 & 710 & $\cdots$ & $\cdots$ & 0.66$^{+0.09}_{-0.09}$ & 10 \\
NGC 4494 & E1 & 16.6 & --24.2 &  1344 & 117 & 1338 & $\cdots$ & $\cdots$ & 0.60$^{+0.12}_{-0.11}$ & 10 \\
\enddata
\tablenotetext{a}{Hubble type from \citet{dev91}.}
\tablenotetext{b}{Distance in units of Mpc: M85, M60, M87, M49 \citep{mei07}, NGC 1407 \citep{spo08}, and other galaxies \citep{ton01}.}
\tablenotetext{c}{$K-$band absolute magnitude from the Two Micron All Sky Survey \citep{skr06} corrected for foreground Galactic extinction \citep{sf11}.}
\tablenotetext{d}{Systemic velocity: M85 (this study), M60 \citep{lee08}, M87, M49 \citep{smi00}, NGC 1399 \citep{ric04}, NGC 5128 \citep{hui95}, NGC 4636 \citep{par10}, NGC 1407 \citep{rom09}, and other galaxies \citep{cap11}.}
\tablenotetext{e}{References from which the GC velocity data are retrieved are in parentheses. (1) \citet{hwa08}, (2) \citet{cot01}, (3) \citet{cot03}, (4) \citet{ric04}, (5) \citet{pff04}, (6) \citet{sch06}, (7) \citet{lee10}, (8) \citet{rom09}, (9) \citet{pot15}, (10) \citet{pot13}, (11) \citet{csb08}}
\end{deluxetable*}

In {\bf Figure \ref{fig:measure_dist2}}, we compare the ages and metallicities of the M85 GCs with those of the Milky Way GCs and M31 GCs that were derived from the full spectrum fitting with ULySS.  
We used the ages and metallicities of the Milky Way GCs and M31 GCs given by \citet{cez13}.
We note that the GCs in M31 and M85 consist of not only old GCs but also intermediate-age GCs (with age $<$ 8 Gyr), while the GCs in the Milky Way Galaxy are predominantly old ($>$ 10 Gyr). 
The intermediate-age GCs in M31 and M85 are, on average, more metal-rich than the old GCs in the same galaxies,  which is consistent with the results for the GCs in other early-type galaxies \citep{puz05, pie06a, par12}.

The intermediate-age GCs in M85 have, on average, an age of 3.7 $\pm$ 1.9 Gyr and [Fe/H] of --0.26 $\pm$ 0.62 dex, while the old GCs have a lower mean metallicity ([Fe/H] = --1.02 $\pm$ 0.56).
The intermediate-age GCs comprise 55\% of the observed GCs in our sample.
The stars in the nucleus of M85 derived in this study have a mean age of 3.8 $\pm$ 0.1 Gyr and a mean metallicity of [Fe/H] = 0.22 dex.
The ages and metallicities of the central stars and the intermediate-age GCs are consistent within their uncertainties.
The mean [Fe/H] of the intermediate-age GCs is lower than the [Fe/H] of the M85 nucleus, but the [Fe/H] distribution of the intermediate-age GCs has a peak near the [Fe/H] of the M85 nucleus with a low-metallicity tail (see {\bf Figure \ref{fig:measure_dist1}}). We conclude that most of the intermediate-age GCs might have been formed during a wet merger about 4 Gyr ago, after the metal enrichment had proceeded. 
However, we caution that the ages and metallicities of the intermediate-age GCs in M85 cover large ranges with large estimation errors.

The GC system of M85 rotates strongly with a rotation amplitude of $\Omega R_{\rm icor} = 235^{+107}_{-66}$ and a rotation parameter $\Omega R_{\rm icor}/\sigma_{\rm r,cor}$ of 1.47$^{+1.05}_{-0.48}$.
To date, the kinematics of the GC systems in dozens of early-type galaxies have been studied.
\citet{hwa08}, \citet{lee10}, and \citet{pot13} presented the kinematics of the GC systems in early-type galaxies, using a method similar to the one used in this study.
{\bf Table \ref{tab:kinematics_etg}} compares the kinematic parameters of the M85 GC system with those of the GC systems in other early-type galaxies from the above references.
The rotation parameters, $\Omega R / \sigma_{r, cor}$, of most early-type galaxies in the previous studies are smaller than 1, while NGC 7457 shows a much larger value, $\Omega R / \sigma_{r, cor}=1.68^{+0.37}_{-0.40}$ \citep{pot13}.
NGC 7457 (S0) is the faintest among the galaxies in {\bf Table \ref{tab:kinematics_etg}}. 
\citet{csb08} investigated the GCs in NGC 7457 using HST photometry and Keck/LRIS spectroscopy, and found that there is an intermediate-age population with an age of 2--2.5 Gyr.
Based on $N-$body simulations of galaxy mergers given by \citet{bjc05}, 
\citet{pot13} pointed out that the strong rotation feature of NGC 7457 GCs could be explained by uneven-mass galaxy mergers with a 1:10 mass ratio.

M85 ($M_K = -25.1$ mag) is much brighter (much more massive) than NGC 7457 ($M_K = -22.4$ mag).
According to the hierarchical galaxy formation scenario, M85 must have undergone many more merging events than NGC 7457.
In consequence, it is expected that the GC system in M85 may be a  pressure-supported system now.
However, the GC system of M85 shows the largest rotation parameter among the early-type galaxies in {\bf Table \ref{tab:kinematics_etg}}.
This indicates that M85 may have experienced an off-center major merging event which injected the angular momentum to the GC system.
\citet{bek05} presented N-body simulation results for the kinematic properties of GC systems after major dry mergers of Milky-Way-like progenitors.
They assumed that the GC systems of the progenitors are only supported by velocity dispersion. 
They found that the GC systems show significant amounts of rotation after a major dry merger with a mass ratio of 1. 
However, the values of the rotation parameters are smaller than 1 (0.2--0.3), and they are much smaller than the value for M85 GCs.
Thus, further simulation studies are needed to understand the peculiar rotation of the M85 GC system.

\section{S\MakeLowercase{ummary}}

We present a spectroscopic study of GC candidates in the $5\farcm5 \times 5\farcm5$ field of the merger remnant galaxy M85 using Gemini-N/GMOS.
We also included the hypercompact cluster M85-HCC1 \citep{san15} and the M85 nucleus for comparison.
Our main results are summarized as follows.

\begin{itemize}

\item Of the 21 GC candidates, 20 are found to be members of M85 based on their radial velocities. One GC candidate turns out to be a foreground star.

\item
Eleven of the GCs are in an intermediate-age population, and the rest are in an old one ($>$ 8 Gyr).
The mean ages of the old and intermediate-age GCs are 13.3 $\pm$ 3.3 Gyr and 3.7 $\pm$ 1.9 Gyr, respectively.
The mean metallicity of the intermediate-age GCs ([Fe/H] = --0.26) is higher than that of the old GCs ([Fe/H] = --1.02).

\item
The mean radial velocity of the GCs in M85 is \vrad = 724 \kmsend, which is similar to the systemic velocity of M85.
The GC system shows a considerable rotation feature, the amplitude of which is $\Omega R_{\rm icor}$ = 235$^{+107}_{-66}$ \kmsend.
The rotation axis of the GC system shows little agreement with that of the central stellar light ($R<1\arcmin$), indicating that the galaxy is the product of merging.

\item
The rotation-corrected radial velocity dispersion of the GC system of M85 is $\sigma_{\rm r,cor} = 160^{+17}_{-29}$ \kmsend.
The rotation parameter $\Omega R_{\rm icor}/\sigma_{\rm r,cor}$ of the GC system is derived to be 1.47$^{+1.05}_{-0.48}$, which is one of the largest among known early-type galaxies.
%which is the largest among known early-type galaxies.
\end{itemize}

Our results suggest that M85 experienced a wet merger about 4 Gyr ago, resulting in the formation of an intermediate-age nucleus. Most of the intermediate-age GCs might have been formed at the same epoch when the stars in the nucleus were formed.
The strong rotation features of the GC system in M85 can be explained by an off-center major merger.

\acknowledgments
We would like to thank anonymous referee for her/his useful comments
on the manuscript and Brian S. Cho for his helpful comments in improving the English of this paper.
This work was supported by the National Research Foundation of Korea (NRF) grant
funded by the Korean Government (NRF-2017R1A2B4004632).
This work was supported by the K-GMT Science Program (PID: GEMINI-KR-2015A-046) of Korea Astronomy and Space Science Institute (KASI).



\begin{thebibliography}{}

\bibitem[Alam et al.(2015)]{ala15} Alam, S., Albareti, F.~D., Allende Prieto, C., et al.\ 2015, \apjs, 219, 12 
\bibitem[Bekki et al.(2005)]{bek05} Bekki, K., Beasley, M.~A., Brodie, J.~P., \& Forbes, D.~A.\ 2005, \mnras, 363, 1211 
\bibitem[Binggeli et al.(1985)]{bst85} Binggeli, B., Sandage, A., \& Tammann, G.~A.\ 1985, \aj, 90, 1681 
\bibitem[Bournaud et al.(2005)]{bjc05} Bournaud, F., Jog, C.~J., \& Combes, F.\ 2005, \aap, 437, 69 
\bibitem[Burstein(1979)]{bur79} Burstein, D.\ 1979, \apjs, 41, 435
\bibitem[Burstein et al.(1984)]{bur84} Burstein, D., Faber, S.~M., Gaskell, C.~M., \& Krumm, N.\ 1984, \apj, 287, 586 
\bibitem[Cappellari et al.(2011)]{cap11} Cappellari, M., Emsellem, E., Krajnovi{\'c}, D., et al.\ 2011, \mnras, 413, 813 
\bibitem[Cezario et al.(2013)]{cez13} Cezario, E., Coelho, P.~R.~T., Alves-Brito, A., Forbes, D.~A., \& Brodie, J.~P.\ 2013, \aap, 549, A60 
\bibitem[Chomiuk et al.(2008)]{csb08} Chomiuk, L., Strader, J., \& Brodie, J.~P.\ 2008, \aj, 136, 234 
\bibitem[Cohen et al.(2003)]{cbc03} Cohen, J.~G., Blakeslee, J.~P., \& C{\^o}t{\'e}, P.\ 2003, \apj, 592, 866 
\bibitem[C{\^o}t{\'e} et al.(2001)]{cot01} C{\^o}t{\'e}, P., McLaughlin, D.~E., Hanes, D.~A., et al.\ 2001, \apj, 559, 828 
\bibitem[C{\^o}t{\'e} et al.(2003)]{cot03} C{\^o}t{\'e}, P., McLaughlin, D.~E., Cohen, J.~G., \& Blakeslee, J.~P.\ 2003, \apj, 591, 850 
\bibitem[C{\^o}t{\'e} et al.(2004)]{cot04} C{\^o}t{\'e}, P., Blakeslee, J.~P., Ferrarese, L., et al.\ 2004, \apjs, 153, 223 
\bibitem[de Freitas Pacheco \& Barbuy(1995)]{db95} de Freitas Pacheco, J.~A., \& Barbuy, B.\ 1995, \aap, 302, 718 
\bibitem[de Vaucouleurs et al.(1991)]{dev91} de Vaucouleurs, G., de Vaucouleurs, A., Corwin, H.~G., Jr., et al.\ 1991, Third Reference Catalogue of Bright Galaxies, Vol. I: Explanations and References, Vol. II: Data for galaxies Between 0$^{h}$ and 12$^{h}$, Vol. III: Data for galaxies Between 12$^{h}$ and 24$^{h}$. (New York, USA: Springer)
\bibitem[Durrell et al.(2014)]{dur14} Durrell, P.~R., C{\^o}t{\'e}, P., Peng, E.~W., et al.\ 2014, \apj, 794, 103 
\bibitem[Emsellem et al.(2007)]{ems07} Emsellem, E., Cappellari, M., Krajnovi{\'c}, D., et al.\ 2007, \mnras, 379, 401 
\bibitem[Emsellem et al.(2011)]{ems11} Emsellem, E., Cappellari, M., Krajnovi{\'c}, D., et al.\ 2011, \mnras, 414, 888 
\bibitem[Ferrarese et al.(2006)]{fer06} Ferrarese, L., C{\^o}t{\'e}, P., Jord{\'a}n, A., et al.\ 2006, \apjs, 164, 334 
\bibitem[Fisher et al.(1996)]{ffi96} Fisher, D., Franx, M., \& Illingworth, G.\ 1996, \apj, 459, 110 
\bibitem[Gavazzi et al.(2004)]{gav04} Gavazzi, G., Zaccardo, A., Sanvito, G., Boselli, A., \& Bonfanti, C.\ 2004, \aap, 417, 499 
\bibitem[Graves \& Schiavon(2008)]{gs08} Graves, G.~J., \& Schiavon, R.~P.\ 2008, \apjs, 177, 446-464 
\bibitem[Hui et al.(1995)]{hui95} Hui, X., Ford, H.~C., Freeman, K.~C., \& Dopita, M.~A.\ 1995, \apj, 449, 592 
\bibitem[Hwang et al.(2008)]{hwa08} Hwang, H.~S., Lee, M.~G., Park, H.~S., et al.\ 2008, \apj, 674, 869
\bibitem[Jord{\'a}n et al.(2009)]{jor09} Jord{\'a}n, A., Peng, E.~W., Blakeslee, J.~P., et al.\ 2009, \apjs, 180, 54 
\bibitem[Koleva et al.(2008)]{kol08} Koleva, M., Prugniel, P., Ocvirk, P., Le Borgne, D., \& Soubiran, C.\ 2008, \mnras, 385, 1998 
\bibitem[Koleva et al.(2009)]{kol09} Koleva, M., Prugniel, P., Bouchard, A., \& Wu, Y.\ 2009, \aap, 501, 1269 
\bibitem[Kormendy et al.(2009)]{kor09} Kormendy, J., Fisher, D.~B., Cornell, M.~E., \& Bender, R.\ 2009, \apjs, 182, 216
\bibitem[Krajnovi{\'c} et al.(2011)]{kra11} Krajnovi{\'c}, D., Emsellem, E., Cappellari, M., et al.\ 2011, \mnras, 414, 2923 
\bibitem[Lauer et al.(2005)]{lau05} Lauer, T.~R., Faber, S.~M., Gebhardt, K., et al.\ 2005, \aj, 129, 2138 
\bibitem[Lee et al.(2008)]{lee08} Lee, M.~G., Park, H.~S., Kim, E., et al.\ 2008, \apj, 682, 135-154 
\bibitem[Lee et al.(2010)]{lee10} Lee, M.~G., Park, H.~S., Hwang, H.~S., et al.\ 2010, \apj, 709, 1083 
\bibitem[Li et al.(2015)]{li15} Li, B., Peng, E.~W., Zhang, H.-x., et al.\ 2015, \apj, 806, 133 
\bibitem[Lim et al.(2017)]{lim17} Lim, S., Peng, E.~W., Duc, P.-A., et al.\ 2017, \apj, 835, 123  
\bibitem[Liu et al.(2013)]{liu13} Liu, G.-C., Lu, Y.-J., Chen, X.-L., Du, W., \& Zhao, Y.-H.\ 2013, Research in Astronomy and Astrophysics, 13, 1025-1040 
%\bibitem[Mar{\'{\i}}n-Franch et al.(2009)]{mar09} Mar{\'{\i}}n-Franch, A., Aparicio, A., Piotto, G., et al.\ 2009, \apj, 694, 1498 
\bibitem[McDermid et al.(2004)]{mcd04} McDermid, R., Emsellem, E., Cappellari, M., et al.\ 2004, Astronomische Nachrichten, 325, 100 
\bibitem[Mei et al.(2007)]{mei07} Mei, S., Blakeslee, J.~P., C{\^o}t{\'e}, P., et al.\ 2007, \apj, 655, 144
\bibitem[Park et al.(2010)]{par10} Park, H.~S., Lee, M.~G., Hwang, H.~S., et al.\ 2010, \apj, 709, 377 
\bibitem[Park et al.(2012)]{par12} Park, H.~S., Lee, M.~G., Hwang, H.~S., et al.\ 2012, \apj, 759, 116 
\bibitem[Peng et al.(2004)]{pff04} Peng, E.~W., Ford, H.~C., \& Freeman, K.~C.\ 2004, \apj, 602, 705 
\bibitem[Peng et al.(2006)]{pen06} Peng, E.~W., Jord{\'a}n, A., C{\^o}t{\'e}, P., et al.\ 2006, \apj, 639, 95
\bibitem[Pierce et al.(2006a)]{pie06a} Pierce, M., Beasley, M.~A., Forbes, D.~A., et al.\ 2006a, \mnras, 366, 1253 
\bibitem[Pierce et al.(2006b)]{pie06b} Pierce, M., Bridges, T., Forbes, D.~A., et al.\ 2006b, \mnras, 368, 325 
\bibitem[Pota et al.(2013)]{pot13} Pota, V., Forbes, D.~A., Romanowsky, A.~J., et al.\ 2013, \mnras, 428, 389 
\bibitem[Pota et al.(2015)]{pot15} Pota, V., Brodie, J.~P., Bridges, T., et al.\ 2015, \mnras, 450, 1962 
\bibitem[Proctor et al.(2004)]{pro04} Proctor, R.~N., Forbes, D.~A., \& Beasley, M.~A.\ 2004, \mnras, 355, 1327 
\bibitem[Puzia et al.(2005)]{puz05} Puzia, T.~H., Kissler-Patig, M., Thomas, D., et al.\ 2005, \aap, 439, 997 
\bibitem[Richtler et al.(2004)]{ric04} Richtler, T., Dirsch, B., Gebhardt, K., et al.\ 2004, \aj, 127, 2094 
\bibitem[Romanowsky et al.(2009)]{rom09} Romanowsky, A.~J., Strader, J., Spitler, L.~R., et al.\ 2009, \aj, 137, 4956 
\bibitem[S{\'a}nchez-Bl{\'a}zquez et al.(2006)]{san06} S{\'a}nchez-Bl{\'a}zquez, P., Peletier, R.~F., Jim{\'e}nez-Vicente, J., et al.\ 2006, \mnras, 371, 703 
\bibitem[Sandoval et al.(2015)]{san15} Sandoval, M.~A., Vo, R.~P., Romanowsky, A.~J., et al.\ 2015, \apjl, 808, L32
\bibitem[Schiavon(2007)]{sch07} Schiavon, R.~P.\ 2007, \apjs, 171, 146 
\bibitem[Schlafly \& Finkbeiner(2011)]{sf11} Schlafly, E.~F., \& Finkbeiner, D.~P.\ 2011, \apj, 737, 103 
\bibitem[Schuberth et al.(2006)]{sch06} Schuberth, Y., Richtler, T., Dirsch, B., et al.\ 2006, \aap, 459, 391 
\bibitem[Schweizer \& Seitzer(1988)]{ss88} Schweizer, F., \& Seitzer, P.\ 1988, \apj, 328, 88 
\bibitem[Schweizer et al.(1990)]{sch90} Schweizer, F., Seitzer, P., Faber, S.~M., et al.\ 1990, \apjl, 364, L33
\bibitem[Schweizer \& Seitzer(1992)]{ss92} Schweizer, F., \& Seitzer, P.\ 1992, \aj, 104, 1039 
\bibitem[Sharina et al.(2010)]{sha10} Sharina, M.~E., Chandar, R., Puzia, T.~H., Goudfrooij, P., \& Davoust, E.\ 2010, \mnras, 405, 839 
\bibitem[Skrutskie et al.(2006)]{skr06} Skrutskie, M.~F., Cutri, R.~M., Stiening, R., et al.\ 2006, \aj, 131, 1163 
\bibitem[Smith et al.(2000)]{smi00} Smith, R.~J., Lucey, J.~R., Hudson, M.~J., Schlegel, D.~J., \& Davies, R.~L.\ 2000, \mnras, 313, 469 
\bibitem[Spolaor et al.(2008)]{spo08} Spolaor, M., Forbes, D.~A., Hau, G.~K.~T., Proctor, R.~N., \& Brough, S.\ 2008, \mnras, 385, 667 
\bibitem[Terlevich \& Forbes(2002)]{tf02} Terlevich, A.~I., \& Forbes, D.~A.\ 2002, \mnras, 330, 547 
\bibitem[Thomas et al.(2003)]{tmb03} Thomas, D., Maraston, C., \& Bender, R.\ 2003, \mnras, 339, 897 
\bibitem[Thomas et al.(2011)]{tmj11} Thomas, D., Maraston, C., \& Johansson, J.\ 2011, \mnras, 412, 2183 
\bibitem[Toloba et al.(2016)]{tol16} Toloba, E., Li, B., Guhathakurta, P., et al.\ 2016, \apj, 822, 51 
\bibitem[Tonry \& Davis(1979)]{td79} Tonry, J., \& Davis, M.\ 1979, \aj, 84, 1511 
\bibitem[Tonry et al.(2001)]{ton01} Tonry, J.~L., Dressler, A., Blakeslee, J.~P., et al.\ 2001, \apj, 546, 681 
\bibitem[Trager et al.(1998)]{tra98} Trager, S.~C., Worthey, G., Faber, S.~M., Burstein, D., \& Gonz{\'a}lez, J.~J.\ 1998, \apjs, 116, 1 
\bibitem[Trancho et al.(2014)]{tra14} Trancho, G., Miller, B.~W., Schweizer, F., Burdett, D.~P., \& Palamara, D.\ 2014, \apj, 790, 122 
\bibitem[Vazdekis et al.(2010)]{vaz10} Vazdekis, A., S{\'a}nchez-Bl{\'a}zquez, P., Falc{\'o}n-Barroso, J., et al.\ 2010, \mnras, 404, 1639 
\bibitem[Worthey et al.(1994)]{wor94} Worthey, G., Faber, S.~M., Gonzalez, J.~J., \& Burstein, D.\ 1994, \apjs, 94, 687 
\bibitem[Worthey \& Ottaviani(1997)]{wo97} Worthey, G., \& Ottaviani, D.~L.\ 1997, \apjs, 111, 377 

\end{thebibliography}
\end{document}